\documentclass[twocolumn,superscriptaddress,a4paper,nopdfoutputerror]{quantumarticle}

\usepackage{amsmath,amssymb,amsfonts,mathtools,amsthm,bm}
\usepackage{graphicx}
\usepackage{tikz}
\usetikzlibrary{arrows.meta,positioning}
\usepackage[numbers,sort&compress]{natbib}
\usepackage{nameref}

\newtheorem{proposition}{Proposition}
\newtheorem{theorem}[proposition]{Theorem}
\newtheorem{lemma}[proposition]{Lemma}
\newtheorem{corollary}[proposition]{Corollary}

\providecommand{\ket}[1]{\lvert#1\rangle}
\providecommand{\bra}[1]{\langle#1\rvert}

\newcommand{\Ran}{\operatorname{Ran}}
\newcommand{\spec}{\operatorname{spec}}

\title{Orbit-resolved spectra and dynamical accessibility in
multi-register covering Hamiltonians}

\author{Fabrício de Souza Luiz}
\orcid{0000-0002-6375-0939}
\email{fsluiz@unicamp.br}
\affiliation{Instituto de Física Gleb Wataghin, Universidade Estadual de Campinas,
13083-859 Campinas, SP, Brazil}

\date{August 27, 2026}

\begin{document}
\maketitle

\begin{abstract}
Which part of a Hamiltonian spectrum is physically relevant when the
initial state and the interpolation preserve several symmetries?  We answer
this question for a multi-register encoding of fixed-cardinality Set Cover.
The joint action combines register permutations with the faithful base
action of the incidence automorphism group.  Its orbits are unlabelled
candidate covers, and the exact orbit quotient is a hopping Hamiltonian on
those candidates.  We prove that the classical penalties are orbit
invariants and distinguish the resulting fixed space from the generally
smaller common invariant cyclic envelope selected by the initial state and
the Hamiltonian family.  For a specified schedule, the actual trajectory
span can be smaller still.  A sector-resolved
Schur-complement bound then certifies cover-measurement probability with the
kinetic floor required by sector projection.  For the retained-walk linear
path, a general hopping-cancellation point makes the spectrum exactly
orbit-resolved; multiplicities outside the fixed sector are dynamically
dark, whereas transverse dark crossings obey a separate stability theorem.
On even cycles the fixed-sector excitation gap at the cancellation point is
$\lambda(n-1)/(2n-1)$, where $\lambda$ is the uncovered-task penalty,
despite the global multiplicity closure.  Separately,
on the same instance family, we construct a different Johnson--Metropolis
parent path on the injective hard-core space, from a Dicke state to a
Gibbs-amplitude state; this second construction provides a cyclic-envelope
gap bound uniformly along an entire path rather than only at a single
cancellation point.  Its
cover-failure probability is bounded by a constant times $n^{-5}$ and its
cyclic-envelope gap is bounded below by $1024\,n^{-13}$ uniformly along the
path.
Quantitative derivative bounds yield a conditional polynomial adiabatic
runtime in the abstract Hamiltonian-access model.  We make no
quantum-speedup claim: the result is a protocol-dependent separation of
global, symmetry-fixed, cyclic-envelope, and tracked spectra.
\end{abstract}

\section{Introduction}
\label{sec:rewrite-intro}

Symmetry block diagonalizes a Hamiltonian, but block diagonalization alone
does not determine which eigenstates a protocol can occupy.  If an initial
state is fixed by every preserved symmetry, evolution is confined to the
joint fixed space.  Even that statement can be loose: the algebra generated
by the endpoint Hamiltonians may act cyclically only on a proper subspace of
the joint fixed space.  This cyclic subspace is a common invariant envelope,
not a claim that every vector in it is reachable under one prescribed
schedule.  A small global gap can then describe a crossing that the protocol
neither couples to nor detects.

Set Cover, whose minimum-cardinality version is NP-hard
\cite{karp1972reducibility}, admits the usual one-binary-variable-per-base Ising formulation,
and it also appears among the constrained examples motivating quantum
alternating-operator ansatzes~\cite{Lucas2014,Hadfield2019}.  Separately,
qubit-efficient variational encodings show that logarithmic or otherwise
compressed register counts are not by themselves a novelty
claim~\cite{TanEncoding2021,GlosKrawiecZimboras2022,Chancellor2019}.  Our
multi-register address encoding belongs to this broad line; no result below
depends on claiming a new qubit count.

Classical objective symmetries can confine quantum approximate optimization
algorithm (QAOA) dynamics and impose equal measurement probabilities on
symmetry-related strings
\cite{ShaydulinHadfieldHoggSafro2021}.  Quantum walks on automorphism
quotients likewise identify a smaller invariant evolution for suitable
initial states~\cite{KroviBrun2007}, while Krylov methods use repeated
generator action to identify a minimal dynamical subspace
\cite{Saad1992}.  In adiabatic computation, symmetry-adapted path design was
studied by Schaller and Sch\"utzhold~\cite{SchallerSchuetzhold2010}, and
symmetry-restricted effective gaps of stoquastic Hamiltonians were connected
to classical simulability by Bringewatt and Jarret
\cite{BringewattJarret2020}.  These results make an effective sector
physically relevant but do not identify the joint fixed space with the
smaller endpoint-algebra envelope considered here.  Conversely, modern
time-dependent Krylov constructions adapt the basis directly to a driven
trajectory~\cite{TakahashiDelCampo2025}.  We use a different, exact object:
the common invariant envelope of the endpoint family.  A specified schedule
traces a potentially smaller subset of it, so no controllability converse is
assumed.

Finally, symmetric discriminants of reversible Markov chains and annealing
paths with Gibbs-amplitude ground states are established tools in quantum
walk and quantum simulated-annealing constructions
\cite{Szegedy2004,SommaBoixoBarnumKnill2008}.  Quantitative adiabatic
computation originates with the interpolation framework of
Farhi~et~al.~\cite{Farhi2000}; modern reviews emphasize that runtime claims
require a specified path and access model~\cite{AlbashLidar2018}.
Quantitative statements require an isolated spectral band together with
derivative and gap control~\cite{JansenRuskaiSeiler2007}.  We use the discriminant only to
construct a rigorously gapped witness path and make no speedup claim.

Minimum Set Cover (MSC) provides a concrete setting in which these three
spaces differ.  With $n_B$ candidate bases, a $k$-register encoding
represents an ordered candidate by $k\lceil\log_2n_B\rceil$ address qubits.
Permuting the registers gives an action of the symmetric group $S_k$, while
colour-preserving automorphisms of the incidence relation permute the base
labels~\cite{McKayPiperno2014}.  Register symmetry alone and a
final-Hamiltonian sector gap are insufficient for an adiabatic or otherwise
path-dependent statement: one must use the faithful joint action, the common
cyclic envelope selected by the initial state, and the minimum isolation gap
of the band tracked along an explicit path.

This work makes that distinction its central question.  Its three main
contributions are:
\begin{enumerate}
\item an orbit description of the $k$-register encoding under
$S_k\times G_B$, where $G_B$ is the faithfully represented group of base
automorphisms, $\mathcal H_{\rm sym}$ is the subspace fixed by this joint
action, and $\mathcal K_u$ is the common invariant cyclic envelope generated
from the uniform initial state by the endpoint algebra; this gives invariant covering
penalties, exact P\'olya--Redfield orbit counts
\cite{Redfield1927,Polya1937}, and a constructive distinction between these
two spaces;
\item a sector-localization proposition with its kinetic floor, together
with a general orbit-resolved hopping-cancellation theorem and a separate
stability theorem for transverse inter-sector crossings; and
\item an even-cycle illustration of the cancellation theorem and,
separately, a parent protocol built from Metropolis rates on Johnson-graph
moves~\cite{Metropolis1953,BrouwerCohenNeumaier1989}, whose cyclic-envelope gap
is uniformly polynomial, yielding a conditional polynomial-time adiabatic
corollary.
\end{enumerate}

The last result concerns a protocol-dependent invariant gap, not computational
advantage.  The controlled cycle family is classically easy, and the proven
exponent is conservative.  Its role is to separate the gap question cleanly
into symmetry, geometry, and kinetics, without inferring asymptotic behavior
from finite-size data.

Neither orbit reduction, P\'olya enumeration, compact addressing, nor a
Markov discriminant is claimed as new in isolation.  The contribution is
the protocol-dependent link between orbit potentials, group-fixed spectra,
cyclic-envelope closure, tracked bands, sector localization, and dark
multiplicity, together with a separately controlled parent path.

We use the following notation throughout.  For an operator $X$,
$\Ran X$ and $\spec X$ denote its range and spectrum.  The symbol
$\|\cdot\|$ denotes the Hilbert-space norm on vectors and the induced
operator norm on operators.  The gap of a finite-dimensional Hermitian
operator is the distance from its lowest eigenspace to the next distinct
eigenvalue; $\lambda_{\min}(X|_{\mathcal L})$ is the lowest eigenvalue of
the compression of $X$ to $\mathcal L$.  For positive $g$, the notation
$f=O(g)$ means $|f|\leq Cg$ for a constant $C>0$ independent of the
asymptotic variable in the stated limit.

\section{Encoding, orbit structure, and the cyclic envelope}
\label{sec:rewrite-framework}

Let $\mathcal T=\{1,\ldots,n_T\}$ be a task universe and let
$\mathcal B_b\subseteq\mathcal T$, $1\leq b\leq n_B$, be the candidate
bases.  Equivalently, $A\in\{0,1\}^{n_B\times n_T}$ is their incidence
matrix, with $A_{bt}=1$ exactly when $t\in\mathcal B_b$.  Minimum Set Cover
asks for
\begin{equation}
 k^\star=\min\bigl\{|S|:S\subseteq[n_B],\quad
 \textstyle\bigcup_{b\in S}\mathcal B_b=\mathcal T\bigr\}.
 \label{eq:msc-definition}
\end{equation}
We assume that the instance is coverable, so $k^\star$ is finite.  Because
adding bases cannot uncover a task, the exact-cardinality feasibility
indicator is
\[
 \begin{aligned}
 \chi_k&:=\mathbf 1[\text{an exact-$k$ cover exists}]\\
 &=\mathbf 1[k\geq k^\star],\qquad 1\leq k\leq n_B.
 \end{aligned}
\]
We fix $1\leq k\leq n_B$ address registers and first work in the physical
address space
\begin{equation}
 \mathcal H_{\rm addr}=(\mathbb C^{n_B})^{\otimes k}.
 \label{eq:addr-space-rewrite}
\end{equation}
The embedding into $2^{\lceil\log_2n_B\rceil}$ levels and its dynamically
disconnected padding block are implementation details.

Write $\ket{\boldsymbol b}=\ket{b_1}\otimes\cdots\otimes\ket{b_k}$ for a
computational-basis tuple.  Its numbers of uncovered tasks and repeated
base pairs are
\begin{align}
 U(\boldsymbol b)
 &=\sum_{t=1}^{n_T}\prod_{r=1}^k(1-A_{b_r,t}),\\
 D(\boldsymbol b)
 &=\sum_{1\leq r<s\leq k}\mathbf 1[b_r=b_s],
\end{align}
respectively, where $\mathbf 1[\cdot]$ is the indicator of the enclosed
condition.  Thus $U(\boldsymbol b)=D(\boldsymbol b)=0$ exactly when a
measurement returns an exact-$k$ cover with distinct bases.  When
$k=k^\star$, every such outcome solves MSC; for other $k$, the Hamiltonian
encodes the fixed-cardinality covering problem rather than minimum
cardinality itself.  In particular, if $k<k^\star$, then $\chi_k=0$ and the
exact-$k$ solution subspace is empty.  This is combinatorial infeasibility,
not a dynamically inaccessible solution or a spectral-gap obstruction.  The
register count fixes $k$ externally; the Hamiltonian contains no internal
flag that determines $k^\star$.  An end-to-end MSC algorithm must therefore
add an outer cardinality search or decision procedure, which is outside the
scope of this fixed-$k$ spectral analysis.

Let $\ket e=\sum_{b=1}^{n_B}\ket b$ and let $I_B$ be the
identity on one address register.  The normalized complete-graph hopping
matrix and its $k$-register lift are
\begin{equation}
 W=\frac{\ket e\!\bra e-I_B}{n_B-1},
 \qquad
 H_{\rm walk}=\sum_{r=1}^k W^{(r)},
 \label{eq:walk-definition}
\end{equation}
where $W^{(r)}$ acts as $W$ on register $r$ and as the identity on all
others.  Define the diagonal penalty operators
\begin{equation}
 H_{\rm cov}=\sum_{\boldsymbol b}U(\boldsymbol b)
 \ket{\boldsymbol b}\!\bra{\boldsymbol b},
 \qquad
 H_{\rm excl}=\sum_{\boldsymbol b}D(\boldsymbol b)
 \ket{\boldsymbol b}\!\bra{\boldsymbol b}.
 \label{eq:penalty-definitions}
\end{equation}
Both sums run over all $n_B^k$ computational-basis tuples.
For positive penalty energies $\lambda$ and $\mu$, the problem Hamiltonian
is
\begin{equation}
 H_{\rm prob}=H_{\rm walk}+\lambda H_{\rm cov}+\mu H_{\rm excl},
 \label{eq:hprob-rewrite}
\end{equation}
so $\lambda U(\boldsymbol b)+\mu D(\boldsymbol b)$ is the total diagonal
penalty of $\ket{\boldsymbol b}$.  Although $H_{\rm walk}$ is deliberately
register local, the full problem Hamiltonian is not a noninteracting
register model: $H_{\rm cov}$ depends jointly on all $k$ addresses, while
$H_{\rm excl}$ couples register pairs.  Thus $H_{\rm walk}$ alone neither
creates entanglement nor makes a hop conditional on the other registers,
whereas the nonseparable penalties can generate inter-register correlations
and the walk redistributes the resulting amplitudes.  A correlated or
constraint-conditioned mixer would define a different protocol, generator
algebra, cyclic envelope, and gap problem; it is not analysed here.

An explicit initial Hamiltonian is
\begin{equation}
 H_{\rm init}=\sum_{r=1}^k\bigl(I_B-\ket{u_B}\!\bra{u_B}\bigr)_r,
 \qquad
 \ket{u_B}=n_B^{-1/2}\sum_{b=1}^{n_B}\ket b,
 \label{eq:hinit-rewrite}
\end{equation}
with unique ground state $\ket u=\ket{u_B}^{\otimes k}$ and initial gap one.
The retained-walk linear interpolation is parameterized by $s\in[0,1]$ as
\begin{equation}
 H(s)=(1-s)H_{\rm init}+sH_{\rm prob}.
 \label{eq:cycle-linear-rewrite}
\end{equation}

\subsection{The faithfully represented group}

Viewing the incidence relation as a bipartite graph whose two vertex classes
have different colours, its colour-preserving automorphisms are the
automorphisms that preserve those classes~\cite{McKayPiperno2014}.  Their
group is
\begin{equation}
 \begin{split}
 G_A=\{(\pi_B,\pi_T)\in {}&S_{n_B}\times S_{n_T}:\\
 &A_{\pi_B(b),\pi_T(t)}=A_{bt}\ \text{for all }b,t\},
 \end{split}
\end{equation}
where $S_d$ denotes the symmetric group on $d$ labels.
Only its base component acts on Eq.~\eqref{eq:addr-space-rewrite}.  We
therefore use the effective group
\begin{equation}
 G_B=\operatorname{pr}_B(G_A)\simeq G_A/\ker\rho,
 \qquad \rho(\pi_B,\pi_T)=\pi_B,
 \label{eq:faithful-base-group}
\end{equation}
where $\operatorname{pr}_B$ projects an automorphism onto its base
permutation and $\ker\rho$ contains the automorphisms acting trivially on
all bases.  Thus $G_B$ is the faithfully represented base group.  The
register and base actions commute, so the
represented symmetry is
\begin{equation}
 G=S_k\times G_B.
\end{equation}
Every term in Eqs.~\eqref{eq:hprob-rewrite} and \eqref{eq:hinit-rewrite}
commutes with $G$.  Consequently, any Hamiltonian $H$ commuting with $G$
has the block structure
\begin{align}
 \mathcal H_{\rm addr}
 &\simeq\bigoplus_{\alpha,\beta}
 V_\alpha^{S_k}\otimes V_\beta^{G_B}\otimes\mathcal M_{\alpha\beta},\\
 H&\simeq\bigoplus_{\alpha,\beta}
 I_{V_\alpha}\otimes I_{V_\beta}\otimes h_{\alpha\beta}.
 \label{eq:commutant-rewrite}
\end{align}
Here $\alpha$ and $\beta$ label irreducible representations of $S_k$ and
$G_B$; the spaces $V_\alpha^{S_k}$ and $V_\beta^{G_B}$ carry those
representations; $\mathcal M_{\alpha\beta}$ is their multiplicity space;
and $h_{\alpha\beta}$ is the reduced action on that space.  The symbols
$I_{V_\alpha}$ and
$I_{V_\beta}$ denote the corresponding identities.  This is the source of
Schur multiplicities~\cite{Serre1977}; an equality of eigenvalues in
different blocks is a separate phenomenon.  The full representation and
the role of the kernel of $G_A$ are detailed in
Appendix~\ref{app:joint-representation}.

\subsection{Symmetry-fixed space, cyclic envelope, and trajectory}

Let $U_\sigma$ permute the $k$ registers according to $\sigma\in S_k$, and
let $V_g$ apply the base permutation $g\in G_B$ in every register.  The
joint fixed projector and its range are
\begin{align}
 \Pi_{\rm sym}
 &=\left(\frac1{k!}\sum_{\sigma\in S_k}U_\sigma\right)
 \left(\frac1{|G_B|}\sum_{g\in G_B}V_g\right),\\
 \mathcal H_{\rm sym}&=\Ran\Pi_{\rm sym}.
 \label{eq:hsym-rewrite}
\end{align}
This is the symmetry-fixed invariant space, not automatically the smallest
invariant arena selected by the initial state.  Let
$\mathfrak A=\operatorname{alg}^*(H_{\rm init},H_{\rm prob},I)$ be the
smallest unital algebra containing the endpoints and closed under adjoints;
here $I$ is the identity on $\mathcal H_{\rm addr}$.  The common invariant
cyclic envelope is
\begin{multline}
 \mathcal K_u=\mathfrak A\ket u\\
 =\operatorname{span}\{X_\ell\cdots X_1\ket u:\
 X_i\in\{H_{\rm init},H_{\rm prob}\},\ \ell\geq0\}.
 \label{eq:ku-rewrite}
\end{multline}
It is the smallest common reducing subspace that contains the initial state.
For a specified schedule $s:[0,T]\to[0,1]$, let $U_s(t,0)$ solve
$i\partial_tU_s(t,0)=H(s(t))U_s(t,0)$ and define the actual trajectory set
\begin{equation}
 \mathcal R_u[s(\cdot)]
 =\{U_s(t,0)\ket u:0\leq t\leq T\}.
 \label{eq:trajectory-set}
\end{equation}
The Dyson expansion and invariance of $\mathcal K_u$ under both endpoints
give, in general,
\begin{equation}
 \operatorname{span}\mathcal R_u[s(\cdot)]
 \subseteq\mathcal K_u\subseteq\mathcal H_{\rm sym}
 \subseteq\mathcal H_{\rm addr}.
 \label{eq:nested-spaces-rewrite}
\end{equation}
Either of the first two inclusions can be strict.  Equality between the
trajectory span and $\mathcal K_u$ is a controllability question and is not
assumed here.  Padding makes the last nontrivial inclusion strict
automatically if one defines the symmetry on the full qubit space.

\begin{figure*}[t]
\centering
\resizebox{0.995\textwidth}{!}{%
\begin{tikzpicture}[font=\small,>=Latex]
  \node[draw,rounded corners,minimum width=6.0cm,minimum height=3.1cm]
        (valid) {};
  \node[draw,rounded corners,fill=blue!5,minimum width=4.7cm,
        minimum height=2.1cm] (sym) at ([yshift=-0.25cm]valid.center) {};
  \node[draw,rounded corners,fill=blue!14,minimum width=2.9cm,
        minimum height=1.05cm] (cyc) at ([yshift=-0.25cm]sym.center) {};
  \node[anchor=north west] at ([xshift=0.12cm,yshift=-0.12cm]valid.north west)
        {\bfseries(a) $\mathcal H_{\rm addr}$};
  \node[anchor=north west] at ([xshift=0.12cm,yshift=-0.12cm]sym.north west)
        {$\mathcal H_{\rm sym}$};
  \node at ([yshift=0.28cm]cyc.center) {$\mathcal K_u$};
  \draw[very thick,blue!70!black,-{Latex[length=1.8mm]}]
        ([xshift=-0.72cm,yshift=-0.10cm]cyc.center)
        .. controls ([yshift=0.08cm]cyc.center) ..
        ([xshift=0.72cm,yshift=-0.10cm]cyc.center);
  \node[anchor=north,font=\scriptsize]
        at ([yshift=-0.12cm]cyc.center) {$\mathcal R_u[s(\cdot)]$};
  \node[align=center,below=0.12cm of valid]
        {symmetry fixes $\mathcal H_{\rm sym}$; the generators select $\mathcal K_u$};

  \begin{scope}[xshift=6.0cm,yshift=-1.90cm]
    \draw[->] (0,0) -- (5.2,0) node[below] {$s$};
    \draw[->] (0,0) -- (0,3.2) node[left] {energy};
    \node[anchor=west] at (0.25,3.35) {\bfseries(b) Dark global crossing};
    \draw[very thick,blue] (0.25,0.55) .. controls (2.0,0.75) and
          (3.5,1.10) .. (4.85,1.25)
          node[right,blue,align=left] {tracked ground\\($\gamma$)};
    \draw[very thick,blue,dashed] (0.25,2.55) .. controls (2.0,2.35) and
          (3.5,2.25) .. (4.85,2.15)
          node[right,blue,align=left] {tracked excitation\\($\gamma$)};
    \draw[very thick,red!75!black] (0.25,2.35) .. controls (1.8,1.65) and
          (3.2,0.75) .. (4.85,0.25)
          node[right,red!75!black,align=left,yshift=0.12cm]
          {dark branch\\($\gamma'$)};
    \fill (3.05,1.00) circle (1.7pt);
    \draw[<->,blue] (2.25,0.86) -- (2.25,2.30)
          node[midway,right,align=left] {$\Delta_{\rm ad}>0$\\inside $\mathcal K_u$};
    \node[align=center,below=0.12cm] at (2.7,0)
          {different irreps do not hybridize};
  \end{scope}
\end{tikzpicture}
}
\caption{Protocol-dependent spectral relevance.  (a) The joint-fixed space
is fixed by symmetry; the cyclic envelope also depends on the initial state
and generated algebra, while a specified schedule traces
$\mathcal R_u[s(\cdot)]$ inside it.  Membership in $\mathcal K_u$ alone does
not imply state reachability.  (b) The labels $\gamma$ and $\gamma'$ denote
inequivalent symmetry sectors.  A branch in the latter can cross the followed
branch and close a global multiplicity gap while the tracked band remains
isolated within the cyclic envelope.  Solid blue is the tracked ground branch,
dashed blue its in-sector excitation, and red an inequivalent dark branch.
Panel (b) depicts the transverse mechanism of
Theorem~\ref{thm:dark-crossing-stability}, not the simultaneous diagonal
cancellation of Theorem~\ref{thm:orbit-resolved-cancellation}; the diagram
is not to scale.}
\label{fig:accessible-spectrum}
\end{figure*}
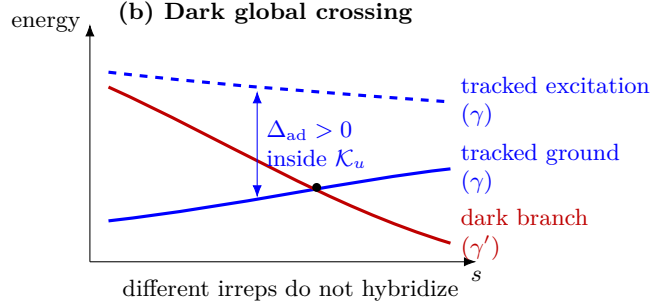

Figure~\ref{fig:accessible-spectrum} schematically summarizes the
distinctions used throughout the paper.  Group invariance identifies a fixed
space, while the initial state and generator algebra select the smaller
common cyclic envelope.  A concrete schedule follows one trajectory inside
it.
An eigenvalue coincidence in another irrep may then close the global
multiplicity gap without closing the isolation gap seen by the protocol.

\subsection{Orbit invariants and exact quotient}

For $\boldsymbol b=(b_1,\ldots,b_k)$, the joint action is
\begin{equation}
 \begin{aligned}
 (\sigma,g)\boldsymbol b
 &=\bigl(g(b_{\sigma^{-1}(1)}),\ldots,
 g(b_{\sigma^{-1}(k)})\bigr),\\
 &\hspace{4.5em}(\sigma,g)\in S_k\times G_B.
 \end{aligned}
 \label{eq:joint-tuple-action}
\end{equation}
The following elementary observation connects this representation directly
to the covering problem.

\begin{lemma}[Orbit invariance of the covering penalties]
\label{lem:orbit-invariance}
The functions $U$ and $D$ are constant on every orbit of the action in
Eq.~\eqref{eq:joint-tuple-action}.  Consequently,
\begin{equation}
 \Phi(\boldsymbol b)=\lambda U(\boldsymbol b)+\mu D(\boldsymbol b)
 \label{eq:orbit-potential}
\end{equation}
is an orbit function, the feasible set $\{U=D=0\}$ is a union of orbits,
and its projector $P=P_k$ commutes with $G$.  It is the zero projector when
$\chi_k=0$.
\end{lemma}

A register-permutation orbit is a size-$k$ multiset of bases.  A joint
$G$-orbit is therefore a $G_B$-equivalence class of such multisets: an
unlabelled candidate $k$-cover, which may still be infeasible or contain
repetitions.  The Cauchy--Frobenius lemma and its P\'olya--Redfield
cycle-index formulation~\cite{Burnside1911,Redfield1927,Polya1937} give
their number as
\begin{proposition}[Orbit classification and count]
\label{prop:orbit-count}
Let $c_\ell(g)$ be the number of length-$\ell$ cycles of $g\in G_B$ in its
base action.  Then
\begin{equation}
 \dim\mathcal H_{\rm sym}
 =\frac1{|G_B|}\sum_{g\in G_B}[z^k]
 \prod_{\ell\geq1}(1-z^\ell)^{-c_\ell(g)}.
 \label{eq:polya-rewrite}
\end{equation}
Here $[z^k]$ extracts the coefficient of $z^k$.  Equivalently, the
right-hand side is the cycle index of the base action evaluated on the
generating functions for multisets.
\end{proposition}

Equation~\eqref{eq:polya-rewrite} counts basis states in the group-fixed
space, not all vectors in the full quantum Hilbert space and not all states
that a particular schedule can prepare.  The cyclic envelope below gives a
rigorous invariant upper bound on the latter set, not a full controllability
characterization.

Orbit quotients of symmetry-confined quantum walks are established
constructions~\cite{KroviBrun2007}; we use the joint action here to retain
the exact Hamiltonian matrix elements.  For an orbit $\mathcal O$, define
\begin{equation}
 \ket{\mathcal O}=|\mathcal O|^{-1/2}
 \sum_{x\in\mathcal O}\ket x.
\end{equation}
These vectors are an orthonormal basis of $\mathcal H_{\rm sym}$.  If $H$
commutes with $G$, $x\in\mathcal O$, and
$H_{yx}:=\langle y|H|x\rangle$, then
\begin{equation}
 \langle\mathcal O'|H|\mathcal O\rangle
 =\sqrt{\frac{|\mathcal O|}{|\mathcal O'|}}
 \sum_{y\in\mathcal O'}H_{yx}.
 \label{eq:orbit-quotient-rewrite}
\end{equation}
The sum is independent of the representative $x$.  Lemma
\ref{lem:orbit-invariance} makes the penalty term diagonal in this basis:
\begin{equation}
 \langle\mathcal O'|V|\mathcal O\rangle
 =\delta_{\mathcal O\mathcal O'}\Phi(\mathcal O).
 \label{eq:orbit-potential-diagonal}
\end{equation}

Thus the quotient is a single-particle hopping problem.  Its sites are
unlabelled candidate covers, $\Phi$ is the on-site potential, and the
off-diagonal entries of Eq.~\eqref{eq:orbit-quotient-rewrite} are hopping
amplitudes with orbit-multiplicity factors.  The graph is connected because
one-register changes connect any two size-$k$ multisets.  When $\chi_k=1$,
reachability of a solution site is therefore not the obstruction: the protocol must
redistribute the state's amplitude into the fixed solution subspace
\begin{equation}
 \Ran P\cap\mathcal H_{\rm sym}
 =\operatorname{span}\{\ket{\mathcal O}:\Phi(\mathcal O)=0\}.
 \label{eq:solution-orbit-span}
\end{equation}
An individual computational solution need not be group fixed, but its
uniform orbit state always lies in Eq.~\eqref{eq:solution-orbit-span}.
Nontrivial combinations within the same solution orbit belong to other
irreps and can be inaccessible from $\ket u$.
When $\chi_k=0$, Eq.~\eqref{eq:solution-orbit-span} is instead the zero
subspace and there is no solution site to reach.
Away from a hopping-cancellation point, orbit potentials are not spectral
gaps because eigenvectors are generally delocalized across several sites.
Moreover, nontrivial irreps have no site in this quotient; this is precisely
why inter-sector partners of dark crossings are absent from the dynamics
shown by the orbit graph.

\subsection{Cyclic-envelope closure in orbit coordinates}

Let
\begin{equation}
 M=\sum_{r=1}^k\bigl(\ket e\!\bra e\bigr)^{(r)},
 \qquad V=\lambda H_{\rm cov}+\mu H_{\rm excl}.
 \label{eq:transport-potential-generators}
\end{equation}
Here the superscript labels the acted-on register:
$\bigl(\ket e\!\bra e\bigr)^{(r)}=
I_B^{\otimes(r-1)}\otimes\ket e\!\bra e\otimes
I_B^{\otimes(k-r)}$.
The endpoint algebra has only these two nontrivial generators.

\begin{lemma}[Two-generator reduction]
\label{lem:two-generator-reduction}
\begin{equation}
 \operatorname{alg}^*(H_{\rm init},H_{\rm prob},I)
 =\operatorname{alg}^*(M,V,I).
 \label{eq:two-generator-reduction}
\end{equation}
\end{lemma}

Let $\widehat M$ and $\widehat V$ be the orbit quotients of $M$ and $V$.
Write $\varphi_1<\cdots<\varphi_p$ for the distinct values of $\Phi$ on
the orbits and let $R_j$ project onto the corresponding potential shell.
The uniform initial coordinate vector $\boldsymbol w$ is defined by its
strictly positive components
\begin{equation}
 (\boldsymbol w)_{\mathcal O}:=\langle\mathcal O|u\rangle
 =\sqrt{|\mathcal O|/n_B^k}>0.
 \label{eq:uniform-orbit-coordinates}
\end{equation}
Thus every orbit has nonzero amplitude initially, but this does not imply
that the individual basis vector $\ket{\mathcal O}$ can be prepared by a
specified schedule.  A strict cyclic inclusion proves that some linear
combinations---patterns of relative amplitudes and phases across orbit
sites---are impossible for every Hamiltonian built from the endpoint
algebra, rather than deleting particular sites from the graph.  Vectors
retained inside $\mathcal K_u$ are only potentially reachable; no converse
is asserted.

\begin{theorem}[Shell-resolved cyclic closure]
\label{thm:shell-cyclic-closure}
In orbit coordinates, $\mathcal K_u$ is the smallest subspace containing
$\boldsymbol w$ and invariant under $\widehat M$ and $\widehat V$.  It is obtained by
the stabilising iteration
\begin{align}
 \mathcal K^{(0)}&=\operatorname{span}\{R_j\boldsymbol w:1\leq j\leq p\},\\
 \mathcal K^{(r+1)}&=\mathcal K^{(r)}+
 \sum_{j=1}^pR_j\widehat M\mathcal K^{(r)}.
 \label{eq:shell-cyclic-iteration}
\end{align}
Moreover, the directions retained by the cyclic envelope and those excluded
from it decompose shell by shell:
\begin{align}
 \mathcal K_u&=\bigoplus_{j=1}^pR_j\mathcal K_u,\\
 \mathcal K_u^\perp\cap\mathcal H_{\rm sym}
 &=\bigoplus_{j=1}^p
 \left(\operatorname{Ran}R_j\ominus R_j\mathcal K_u\right).
 \label{eq:shell-complement-decomposition}
\end{align}
Consequently,
\begin{equation}
 \dim\mathcal K_u\geq p,
 \label{eq:shell-lower-bound}
\end{equation}
and $\mathcal K_u=\mathcal H_{\rm sym}$ whenever $\Phi$ is injective on
orbits.  Conversely, a strict inclusion
$\mathcal K_u\subsetneq\mathcal H_{\rm sym}$ requires two distinct orbits
to share a potential value.

For the linear path in Eq.~\eqref{eq:cycle-linear-rewrite}, let
\begin{equation}
 c(s)=\frac{s}{n_B-1}-\frac{1-s}{n_B}.
 \label{eq:hopping-coefficient}
\end{equation}
Whenever $c(s)\leq0$, the symmetric and cyclic ground energies agree:
\begin{equation}
 E_0^{\mathcal K}(s)=E_0^{\rm sym}(s).
 \label{eq:zero-cyclic-cost-before-cancellation}
\end{equation}
\end{theorem}

The last statement follows physically from the sign-free, connected orbit
hopping problem.  Before cancellation, Perron--Frobenius~\cite{HornJohnson2012}
gives a unique
strictly positive symmetric ground vector, which has nonzero overlap with
$\boldsymbol w$ and hence belongs to the cyclic envelope.  At cancellation, a
nonzero projection of $\boldsymbol w$ remains in the lowest potential shell.  Strict dimension
reduction therefore need not incur a ground-energy cost; the numerical
audit in Section~\ref{sec:numerical-reconstruction} verifies the theorem on
the path grid through $s_c$.  At the problem endpoint $s=1$, outside the
theorem's sign regime, four of the five strict inclusions still have zero
cyclic cost empirically.  Proofs of the orbit
statements and Theorem~\ref{thm:shell-cyclic-closure} are given in
Appendix~\ref{app:joint-representation}.

\section{Sector localization and relevant gaps}
\label{sec:rewrite-localization}

Let $P=P_k$ project onto the span of computational-basis tuples that use
distinct bases and cover every task, and put $Q=I-P$.  By
Lemma~\ref{lem:orbit-invariance}, both projectors commute with the joint
group.  For a normalized state $\psi$, define the exact-cardinality cover
success probability
\begin{equation}
 p_{{\rm cov},k}(\psi)=\langle\psi|P_k|\psi\rangle.
 \label{eq:fixed-k-cover-probability}
\end{equation}
Only when $k=k^\star$ do we write this operational quantity as
$p_{\rm opt}(\psi)$, because then every feasible exact-$k$ outcome is a
minimum cover.  Write
$V=\lambda H_{\rm cov}+\mu H_{\rm excl}$,
$m=\min\{\lambda,\mu\}$ for the minimum single-constraint penalty, and
$\kappa=k/(n_B-1)$ for the magnitude of the lowest possible $k$-register
walk energy.  Here $k$ is the register count fixed in
Section~\ref{sec:rewrite-framework}.  Then $VP=0$ and the complete-graph
normalization gives
\begin{equation}
 QH_{\rm prob}Q\geq(m-\kappa)Q.
 \label{eq:q-lower-rewrite}
\end{equation}

The global localization statement cannot be copied verbatim into an
arbitrary sector.  A projected cover is a superposition of computational
covers and can have nonzero kinetic expectation.  The missing term is the
sector's kinetic floor.

\begin{proposition}[Sector localization]
\label{prop:sector-localization-rewrite}
Let $\gamma=(\alpha,\beta)$ label an isotypic component in the joint
decomposition of Eq.~\eqref{eq:commutant-rewrite}, or more generally let
$\Pi_\gamma$ be any reducing sector projector commuting with
$H_{\rm prob}$, $H_{\rm walk}$, $V$, and $P$.  Suppose
$P_\gamma=P\Pi_\gamma\ne0$.  Define the invalid-sector projector
$Q_\gamma$, the feasible-to-invalid coupling $B_\gamma$, and the sector
kinetic floor $\eta_\gamma$ by
\begin{align}
 Q_\gamma&=Q\Pi_\gamma,\quad\quad
 B_\gamma=Q_\gamma H_{\rm walk}P_\gamma,\\
 \eta_\gamma&=\lambda_{\min}\!\left(
 P_\gamma H_{\rm walk}P_\gamma|_{\Ran P_\gamma}\right).
\end{align}
If $Q_\gamma=0$, every state in the sector lies in $\Ran P$ and the leakage
is identically zero.  Otherwise, let
$E_0^\gamma=\lambda_{\min}(H_{\rm prob}|_{\Ran\Pi_\gamma})$ be the lowest
problem-Hamiltonian energy in the sector and define the compressed
invalid-state separation
\begin{equation}
 \Gamma_\gamma=\lambda_{\min}\!\left(
 Q_\gamma H_{\rm prob}Q_\gamma|_{\Ran Q_\gamma}\right)-E_0^\gamma.
\end{equation}
If $\Gamma_\gamma>0$, every normalized sector ground state $\psi_\gamma$
satisfies
\begin{equation}
 1-\langle\psi_\gamma|P|\psi_\gamma\rangle
 \leq\frac{\|B_\gamma\|^2}
 {\Gamma_\gamma^2+\|B_\gamma\|^2}.
 \label{eq:sector-bound-rewrite}
\end{equation}
Moreover,
\begin{equation}
 \Gamma_\gamma\geq m-\kappa-\eta_\gamma,
 \label{eq:sector-separation-rewrite}
\end{equation}
so the right-hand side of Eq.~\eqref{eq:sector-bound-rewrite} remains valid
with $\Gamma_\gamma$ replaced by
$g_\gamma=m-\kappa-\eta_\gamma>0$.
\end{proposition}

Physically, the certificate compares the energetic separation
$\Gamma_\gamma$ protecting the feasible sector with the coupling
$\|B_\gamma\|$ that leaks amplitude into invalid states.  The kinetic floor
$\eta_\gamma$ is required because a sector-projected cover is a
superposition and need not have zero hopping expectation.  We do not claim
that it changes the numerical certificate on every instance.
The full proof and the cyclic-envelope caveat are in
Appendix~\ref{app:sector-localization}.

The archived audit evaluates every term directly in symmetry sectors of the
full problem Hamiltonian.  For example, in the
$S_3$-trivial $\times D_4$-$E$ sector of \texttt{grid-2x4},
\begin{equation}
 \eta_\gamma=-\frac{2}{7},\quad
 \Gamma_\gamma=4.88189,\quad
 \|B_\gamma\|=0.889694.
 \label{eq:sector-audit-example}
\end{equation}
The worst ground-manifold leakage is $0.0035743$, below both the exact bound
$0.032145$ and the kinetic-floor bound $0.032463$.  Thus the sector kinetic
term is numerically nonzero and changes the certified separation from the
unjustified naive quantity $m-\kappa=32/7$ to
$g_\gamma=34/7$.  Appendix~\ref{app:numerical-protocol} reports all five
$D_4$ sectors and an independent asymmetric instance.

This statement certifies the solution content of the lowest-energy state
within a sector.  It does not say that the ground vector has exact support on
$\Ran P$: for finite penalties Eq.~\eqref{eq:sector-bound-rewrite} controls,
rather than eliminates, leakage.  Exact support occurs only in an infinite
penalty limit or in special decoupled cases.

The proposition applies to genuine commuting symmetry sectors.  A cyclic
envelope projector need not commute with $P$, so its solution probability is
evaluated directly or through a separately constructed commuting projector.

\subsection{Protocol-dependent gaps and initial-state dependence}
\label{sec:rewrite-gaps}

For an invariant space $\mathcal L$, let $P_0^{\mathcal L}(s)$ be the
ground projector of $H(s)|_{\mathcal L}$ and
$g_0^{\mathcal L}(s)=\operatorname{rank}P_0^{\mathcal L}(s)$.  Write the
eigenvalues of the restriction, ordered with multiplicity, as
$E_0^{\mathcal L}(s)\leq E_1^{\mathcal L}(s)\leq\cdots$.  Whenever the
indicated complementary level exists, we keep separate
\begin{align}
 \delta_{\rm mult}^{\mathcal L}(s)
 &=E_1^{\mathcal L}(s)-E_0^{\mathcal L}(s),\\
 \Delta_{\rm exc}^{\mathcal L}(s)
 &=E_{g_0^{\mathcal L}(s)}^{\mathcal L}(s)-E_0^{\mathcal L}(s).
 \label{eq:two-gaps-rewrite}
\end{align}
If $\dim\mathcal L=1$, we set $\delta_{\rm mult}^{\mathcal L}=+\infty$;
if $g_0^{\mathcal L}=\dim\mathcal L$, we set
$\Delta_{\rm exc}^{\mathcal L}=+\infty$.  These conventions express that
the corresponding complementary level does not exist.
The first vanishes whenever the ground state is degenerate; the second is
the excitation gap above the whole manifold.  A Schur multiplicity or an
exact coincidence between irreps is therefore not called a gap closure when
Eq.~\eqref{eq:two-gaps-rewrite}'s second quantity stays positive.

Let $E_0$, $E_0^{\rm sym}$, and $E_0^{\mathcal K}$ be the ground energies
of $H(s)$ on $\mathcal H_{\rm addr}$, $\mathcal H_{\rm sym}$, and
$\mathcal K_u$, respectively, at a common value of $s$.  The nested
invariant spaces in Eq.~\eqref{eq:nested-spaces-rewrite} define the
restriction costs
\begin{align}
 \delta_{\rm sym}&=E_0^{\rm sym}-E_0,\quad\quad
 \delta_{\rm cyc}=E_0^{\mathcal K}-E_0^{\rm sym},\\
 \delta_{\rm acc}&=E_0^{\mathcal K}-E_0
 =\delta_{\rm sym}+\delta_{\rm cyc}.
 \label{eq:accessibility-cost-rewrite}
\end{align}
A positive $\delta_{\rm acc}$ proves that the common cyclic envelope excludes
the global ground branch.  If an isolated band is then followed inside that
envelope, it is globally excited but can nevertheless have high cover
probability.  Conversely, zero restriction cost does not prove that a
particular schedule prepares the corresponding ground vector.

For adiabatic evolution, the relevant object is an isolated spectral band,
not merely the first distinct final eigenvalue
\cite{JansenRuskaiSeiler2007}.  Let
$P_{\rm track}(s)$ be a continuous isolated spectral projector on
$\mathcal K_u$ whose range contains the initial ground state.  Set
$H_{\mathcal K_u}(s)=H(s)|_{\mathcal K_u}$ and let
$H_{\rm track}(s)$ be the compression of $H_{\mathcal K_u}(s)$ to
$\Ran P_{\rm track}(s)$.  Its complementary compression is
\begin{equation}
 H_\perp(s)=H_{\mathcal K_u}(s)
 \big|_{\Ran(I_{\mathcal K_u}-P_{\rm track}(s))},
\end{equation}
where $I_{\mathcal K_u}$ is the identity on $\mathcal K_u$.  For finite
real sets $A,B$, write
$\operatorname{dist}(A,B)=\min_{a\in A,b\in B}|a-b|$ and adopt the
convention $\operatorname{dist}(A,\varnothing)=+\infty$.  Define
\begin{multline}
 \Delta_{\rm ad}(s)=\operatorname{dist}\!\left(
 \spec H_{\rm track}(s),\right.\\
 \left.\spec H_\perp(s)\right).
 \label{eq:adiabatic-gap-rewrite}
\end{multline}
This projector-based definition retains multiplicity: if a complementary
eigenvector has the same numerical eigenvalue as the tracked band, then
$\Delta_{\rm ad}=0$.
An isolated continuous projector has constant rank.  If its intended rank
changes, an isolation gap has closed and a nondegenerate adiabatic formula
cannot simply be continued through that point.

The four quantities have the following physical meanings:
\begin{center}
\centering
\small
\begin{tabular}{lp{0.62\columnwidth}}
\hline
quantity & physical meaning\\
\hline
$\delta_{\rm mult}$ & splitting inside the lowest manifold; zero for a
degenerate ground energy\\
$\Delta_{\rm exc}$ & excitation gap above the entire lowest manifold\\
$\delta_{\rm acc}$ & energy cost of confinement to the common cyclic
envelope\\
$\Delta_{\rm ad}$ & isolation of the spectral band followed inside that
space\\
\hline
\end{tabular}
\end{center}

Both the envelope and the tracked band are conditional on the protocol.  If
$[H(t),\Pi_\gamma]=0$ and the initial state lies in $\Ran\Pi_\gamma$, its
evolution remains there, but an initial state adapted to another irrep or a
symmetry-breaking generator changes the invariant and tracked spectra.  This statement
does not forbid transient concentration in a computationally marked
subspace: it forbids only transition amplitude into symmetry sectors
orthogonal to the one selected by the initial state.

\section{Dark crossings in exact symmetry quotients}
\label{sec:rewrite-crossings}

If $\Pi_a$ and $\Pi_b$ project onto inequivalent irreps of a symmetry
preserved by the entire path, then
\begin{equation}
 \Pi_a\,\partial_sH(s)\,\Pi_b=0.
 \label{eq:dark-coupling-rewrite}
\end{equation}
An equality of the two branch energies is therefore a global crossing but
not an avoided crossing seen by a state confined to either block.

Two distinct mechanisms must be separated.  First, the chosen retained-walk
interpolation has an exactly diagonal point at which register-permutation
multiplicities appear simultaneously.  Second, an isolated pair of
transverse branches can cross between inequivalent irreps.  Only the latter
mechanism is covered by the stability theorem below.

\subsection{Orbit-resolved spectrum at hopping cancellation}

The identities in Eq.~\eqref{eq:app-two-generator-identities} give
\begin{align}
 H(s)&=a(s)I+c(s)M+sV,\\
 a(s)&=k\left(1-s-\frac{s}{n_B-1}\right),
 \label{eq:path-transport-decomposition}
\end{align}
where $c(s)$ is defined in Eq.~\eqref{eq:hopping-coefficient}.  Its unique
zero is
\begin{equation}
 s_c=\frac{n_B-1}{2n_B-1}.
 \label{eq:cycle-cancel-rewrite}
\end{equation}

\begin{theorem}[Orbit-resolved cancellation spectrum]
\label{thm:orbit-resolved-cancellation}
At $s=s_c$,
\begin{equation}
 H(s_c)=s_c(kI+V).
 \label{eq:cancellation-identity}
\end{equation}
Let $\varphi_1<\cdots<\varphi_p$ be the distinct orbit-potential values.
Then the distinct spectra on $\mathcal H_{\rm addr}$,
$\mathcal H_{\rm sym}$, and the cyclic envelope $\mathcal K_u$ are all
\begin{equation}
 \{s_c(k+\varphi_j):1\leq j\leq p\}.
 \label{eq:cancellation-levels}
\end{equation}
In particular, their ground energies and excitation gaps agree at $s_c$;
if $p>1$,
\begin{equation}
 \Delta_{\rm exc}(s_c)=s_c(\varphi_2-\varphi_1).
 \label{eq:cancellation-gap}
\end{equation}
For $p=1$, we adopt the convention $\Delta_{\rm exc}(s_c)=+\infty$.

Assume that at least one exact-$k$ cover with distinct bases exists.  Let
$N_F$ be the number of such base subsets and let $N_{\rm orb}$ be the
number of their $G_B$-orbits.  Then $\varphi_1=0$ and
\begin{align}
 g_0^{\rm addr}(s_c)&=k!N_F,&
 g_0^{\rm sym}(s_c)&=N_{\rm orb},
 \label{eq:cancellation-full-ranks}\\
 1&\leq g_0^{\mathcal K}(s_c)\leq N_{\rm orb}.
 \label{eq:cancellation-cyclic-rank}
\end{align}
When $k=k^\star$, these are minimum-cover counts; otherwise they count
feasible exact-$k$ covers.
\end{theorem}

At cancellation, and only there along this path, the potential separation
between orbit sites is exactly a spectral separation.  For $s<s_c$ the full
Hamiltonian is an irreducible stoquastic matrix~\cite{BravyiEtAl2008Stoquastic}
and has a unique ground
state, whereas Eq.~\eqref{eq:cancellation-full-ranks} is degenerate for every
feasible instance with $k\geq2$.  The resulting global multiplicity closure
contains nontrivial $S_k$ components that are dark to the uniform state.
The symmetric ground space can itself be degenerate when
$N_{\rm orb}>1$, so only a cover-transitive instance
($N_{\rm orb}=1$) has a unique fixed-sector solution orbit.

The interior cancellation is a feature of the endpoint choice
$H_{\rm prob}=H_{\rm walk}+V$.  We retain the walk because this endpoint is
a cost-dressed continuous-time search Hamiltonian rather than a purely
classical cost operator.  For the conventional diagonal endpoint $V$, the
path $(1-s)H_{\rm init}+sV$ has hopping coefficient $-(1-s)/n_B$, which
vanishes only at $s=1$.  The orbit-resolved diagonal spectrum and its
permutation multiplicities still occur at that endpoint, but there is no
interior cancellation point.  We therefore do not present the closure as a
path-independent property of adiabatic Set Cover.

\subsection{Transverse dark crossings}

Unlike the simultaneous diagonal closure, a transverse inter-sector
crossing is locally stable within the symmetry-preserving class.  We use
standard differentiable perturbation theory for isolated Hermitian
branches~\cite{Kato1995}.

\begin{theorem}[Stability of a transverse dark crossing]
\label{thm:dark-crossing-stability}
Let $H_\varepsilon(s)=H(s)+\varepsilon R(s)$ be a finite-dimensional
Hermitian family that is continuously differentiable jointly in
$(s,\varepsilon)$, where $\varepsilon\in\mathbb R$ is
the perturbation strength and $R(s)$ is Hermitian.  Let $\Pi_a,\Pi_b$
project onto
inequivalent symmetry sectors.  Suppose every $H_\varepsilon(s)$ commutes
with both projectors.  Assume that near $(s_\star,0)$ the two restricted
blocks have isolated differentiable reduced eigenvalue branches
$E_a(s,\varepsilon)$ and $E_b(s,\varepsilon)$, simple on their multiplicity
spaces but possibly carrying symmetry-enforced Schur multiplicity, such that
\begin{equation}
 E_a(s_\star,0)=E_b(s_\star,0),
 \qquad
 \partial_s(E_a-E_b)(s_\star,0)\ne0.
 \label{eq:transverse-dark-hypothesis}
\end{equation}
Then, for all sufficiently small $|\varepsilon|$, there is a unique nearby
$s_\star(\varepsilon)=s_\star+O(|\varepsilon|)$ at which the two branches
cross exactly.
If the $a$ branch is separated from the rest of its block by at least
$\Delta_\star>0$ throughout a neighbourhood of $s_\star$ at
$\varepsilon=0$, then its isolation gap remains at least
$\Delta_\star/2$ whenever
$|\varepsilon|\sup_{s\in[0,1]}\|R(s)\|\leq\Delta_\star/4$ and the perturbed crossing
stays in that neighbourhood.
\end{theorem}

Physically, the preserved symmetry forbids the two branches from
hybridizing.  A symmetry-preserving perturbation may move the crossing, but
cannot open an avoided crossing between the sectors.  The complete
stability argument is in Appendix~\ref{app:dark-stability}.

\subsection{A finite joint-symmetry audit}

For the frozen eight-base incidence instance generated from a $2\times4$
vertex grid and specified explicitly in Appendix~\ref{app:d4-catalyst}, take
$k=3$, $\lambda=5$, and $\mu=10$.  Its effective incidence group is
$G_B\simeq D_4$, where $D_4$ is the order-eight dihedral group, and the
$S_3\times D_4$ fixed quotient has dimension $22$,
compared with $512$ in the ordered address space.  Cyclic closure from
$\ket u$ also has dimension $22$, as determined independently by the
generator algebra.  The uncatalysed path has a fixed-space rank closure at
$s=7/15$, showing that joint symmetry alone does not guarantee a rank-one
tracked band or a rank-one cyclic ground manifold.

As a finite illustration of a dark crossing away from the cancellation
point, use the single-register walk $W$ from
Eq.~\eqref{eq:walk-definition} and define
\begin{equation}
 C_2=-\sum_{1\leq r<t\leq k}W^{(r)}W^{(t)},
 \label{eq:d4-catalyst-operator}
\end{equation}
and use the endpoint-preserving path
\begin{equation}
 H_\chi(s)=H(s)+4\chi s(1-s)C_2,
 \label{eq:d4-catalyst-path}
\end{equation}
where $\chi\in\mathbb R$ is the dimensionless catalyst strength.
The operator $C_2$ commutes with both $S_k$ and $G_B$, and its envelope
vanishes at $s=0,1$.  At $\chi=2$, a crossing between different $D_4$
irreps occurs at $s\simeq0.806454$, while the $D_4$-trivial gap there is
approximately $0.258$.  The reduced-branch slope difference is
approximately $-4.54$, so the crossing is transverse.  This diagnostic is
not used in the asymptotic theorem.  Appendix~\ref{app:d4-catalyst}
specifies its numerical protocol,
minimum-gap scan, and residuals.

\subsection{Even cycles as a cover-transitive illustration}

Express minimum vertex cover on the even cycle $C_n$, $n=2k$, as set cover:
bases are vertices, tasks are edges, and each base covers its two incident
edges.  The only minimum covers are the two alternating subsets.  They are
exchanged by a one-site rotation and hence form one orbit of the dihedral
group $D_n$.  Thus this family is cover-transitive, with
$N_F=2$ and $N_{\rm orb}=1$.

Assume
\begin{equation}
 \mu\geq\lambda>0.
 \label{eq:cycle-penalty-hypothesis}
\end{equation}
Any repeated $k$-tuple uses fewer than $k=k^\star$ distinct bases and
therefore leaves at least one edge uncovered; its penalty is at least
$\lambda+\mu$.  A distinct noncover has penalty at least $\lambda$, and
this value is attained by
\begin{equation}
 S_\star=\{1\}\cup\{2j:1\leq j\leq k-1\},
\end{equation}
which leaves exactly one edge uncovered.  Theorem
\ref{thm:orbit-resolved-cancellation} therefore gives
\begin{align}
 s_c&=\frac{n-1}{2n-1},&
 g_0^{\rm addr}(s_c)&=2\cdot k!,\\
 g_0^{\rm sym}(s_c)&=g_0^{\mathcal K}(s_c)=1,&
 \Delta_{\rm exc}^{\mathcal H_{\rm sym}}(s_c)
 &=\lambda\frac{n-1}{2n-1}.
 \label{eq:cycle-dark-closure-rewrite}
\end{align}
The fixed-sector excitation gap therefore increases monotonically from
$3\lambda/7$ at $n=4$ towards $\lambda/2$ as $n\to\infty$, without
attaining that limit, while the global ground
multiplicity becomes $2\cdot k!$.  This is a diagonal,
permutation-induced multiplicity closure, not a transverse two-branch
crossing; Theorem~\ref{thm:dark-crossing-stability} does not assert its
stability under a general symmetry-preserving perturbation.

Equation~\eqref{eq:cycle-dark-closure-rewrite} is pointwise and does not
bound the minimum gap along the retained-walk linear interpolation.  The next
section studies a different parent path on the same family.  Its endpoint,
initial state, and generator algebra are distinct from those above.


\section{A separate parent path and its cyclic-envelope gap}
\label{sec:analytic-family}

We now change protocols while retaining the same even-cycle covering
instances.  The parent Hamiltonians below do not interpolate between
$H_{\rm init}$ and $H_{\rm prob}$: their state space, initial state,
endpoint, and generator algebra are specified independently.  This second
construction controls a gap along an entire path rather than the single
cancellation point of Section~\ref{sec:rewrite-crossings}.  Its purpose is a
gap certificate, not a claim that vertex cover on a cycle is classically
difficult.

\begin{table*}[t]
\centering
\caption{The two protocols share an instance family and orbital language but
not a state space, initial state, endpoint, or complexity statement.}
\label{tab:protocol-comparison}
\small
\setlength{\tabcolsep}{6pt}
\begin{tabular}{p{0.18\textwidth}p{0.36\textwidth}p{0.36\textwidth}}
\hline
feature & retained-walk protocol & Johnson--Metropolis parent protocol\\
\hline
state space & ordered $k$-register address space
$\mathcal H_{\rm addr}$ & injective symmetric hard-core space spanned by
$k$-subsets\\
initial state & product-uniform $\ket u=\ket{u_B}^{\otimes k}$ & Dicke state
$\ket{J_{n,k}}$\\
path & $(1-s)H_{\rm init}+sH_{\rm prob}$ & discriminants $H_{a(s)}$ with
$a(s)=7s$\\
endpoint & retained walk plus covering and exclusion penalties,
$H_{\rm prob}=H_{\rm walk}+V$ & frustration-free Gibbs-amplitude parent
$H_7$\\
common envelope & $\operatorname{alg}^*(H_{\rm init},H_{\rm prob},I)\ket u$
& $\operatorname{alg}^*(\{H_a:0\leq a\leq7\},I)\ket{J_{n,k}}$\\
proved control & exact orbit spectrum at one cancellation point; sector
localisation and dark-branch diagnostics & uniform cyclic-envelope gap,
endpoint concentration, and conditional abstract adiabatic runtime\\
implementation claim & none & none; Dicke preparation and Hamiltonian
simulation are excluded\\
\hline
\end{tabular}
\end{table*}

\subsection{Even-cycle set cover and its joint symmetry}

Let the bases be the vertices of the even cycle $C_n$, $n=2k$, and let the
tasks be its edges.  Base $b$ covers the two edges incident on $b$.  A
$k$-subset $S\subset\mathbb Z_n$, where $\mathbb Z_n$ denotes the vertex
labels modulo $n$, is therefore a cover precisely when it is
one of the two alternating subsets.  The two solutions are exchanged by a
one-site rotation and form one orbit of the dihedral incidence automorphism
group $D_n$, generated by cycle rotations and reflections.  Throughout,
$D_n$ denotes the dihedral group of order $2n$ acting on the $n$ cycle
vertices.

We work in the injective, register-symmetric subspace, with basis
$\{\ket S:|S|=k\}$.  Equivalently, this is the symmetric Dicke embedding of
the hard-core sector of the original registers.  Define
\begin{equation}
 U(S)=|\{i\in\mathbb Z_n:i\notin S,\ i+1\notin S\}|.
 \label{eq:cycle-energy}
\end{equation}
At half filling, $U(S)$ is also the number of adjacent occupied pairs.  Thus
$U=0$ exactly on the two minimum covers.

For a dimensionless annealing parameter $a\geq0$, consider the
continuous-time Metropolis generator~\cite{Metropolis1953} with a
Johnson-graph proposal~\cite{BrouwerCohenNeumaier1989}.  The
Johnson graph $J(n,k)$ has the $k$-subsets of $[n]$ as vertices and joins
$S$ to $T$ exactly when $|S\mathbin\triangle T|=2$, i.e., one selected
vertex is replaced.  For distinct adjacent $S,T$, the transition rate is
\begin{equation}
 q_a(S,T)=\frac1n
 \min\{1,n^{-a[U(T)-U(S)]}\},
 \qquad |S\mathbin\triangle T|=2,
 \label{eq:johnson-metropolis}
\end{equation}
and $q_a(S,T)=0$ otherwise.  Its reversible measure is
\begin{equation}
 \pi_a(S)=Z_a^{-1}n^{-aU(S)}.
 \label{eq:cycle-gibbs}
\end{equation}
Here $Z_a=\sum_{|S|=k}n^{-aU(S)}$ normalizes the probability distribution.
The symmetric-discriminant construction is standard for reversible Markov
chains and Gibbs-amplitude ground states
\cite{Szegedy2004,SommaBoixoBarnumKnill2008}.  Here $H_a$ has diagonal
entry $\langle S|H_a|S\rangle=\sum_{T\ne S}q_a(S,T)$ and, for
$|S\mathbin\triangle T|=2$, off-diagonal entries
\begin{align}
 \langle T|H_a|S\rangle
 &=-\sqrt{q_a(S,T)q_a(T,S)}\\
 &=-\frac1n n^{-a|U(T)-U(S)|/2}.
 \label{eq:cycle-parent}
\end{align}
All remaining off-diagonal entries are zero, so these equations define
$H_a$ completely.
It is frustration free and has the known ground state
\begin{equation}
 \ket{\psi_a}=Z_a^{-1/2}\sum_{|S|=k}n^{-aU(S)/2}\ket S.
 \label{eq:cycle-parent-ground}
\end{equation}
At $a=0$, this is the Dicke state
$\ket{J_{n,k}}=\binom nk^{-1/2}\sum_{|S|=k}\ket S$
\cite{BartschiEidenbenz2019}, and the parent
Hamiltonian $H_{a=0}$ is exactly the Johnson-graph Laplacian at rate $1/n$.
It should not be confused with the restriction of $H_{\rm init}$ from
Eq.~\eqref{eq:hinit-rewrite}, which agrees with a Johnson Laplacian only up
to an additive constant after projection to the injective subspace.  The
whole parent path commutes with $D_n$.

This protocol has its own generator algebra and common cyclic envelope,
distinct from Eq.~\eqref{eq:ku-rewrite}:
\begin{align}
 \mathfrak A_{\rm par}
 &=\operatorname{alg}^{*}\!\left(\{H_a:0\leq a\leq7\}\cup\{I\}\right),\\
 \mathcal K_J^{\rm par}
 &=\mathfrak A_{\rm par}\ket{J_{n,k}}.
 \label{eq:parent-cyclic-space}
\end{align}
Here $I$ is the identity on the half-filled hard-core space.
Let $\mathcal H^{D_n}$ denote the $D_n$-fixed subspace of the half-filled
hard-core space.  Define the fixed-space and cyclic gaps by
\begin{align}
 \Delta_{D_n}(a)
 &=\operatorname{gap}\!\left(H_a|_{\mathcal H^{D_n}}\right),\\
 \Delta_{\rm par}^{\rm cyc}(a)
 &=\operatorname{gap}\!\left(H_a|_{\mathcal K_J^{\rm par}}\right).
 \label{eq:parent-cyclic-gap}
\end{align}

\begin{proposition}[Cyclic envelope of the parent path]
\label{prop:parent-cyclic-accessibility}
For every even $n=2k\geq6$ and $0\leq a\leq7$,
\begin{enumerate}
 \item $\mathcal K_J^{\rm par}$ is a common reducing subspace for the
 family $\{H_a\}_{0\leq a\leq7}$ and
 $\mathcal K_J^{\rm par}\subseteq\mathcal H^{D_n}$;
 \item the unique ground state $\ket{\psi_a}$ belongs to
 $\mathcal K_J^{\rm par}$; and
 \item
 \begin{equation}
  \Delta_{\rm par}^{\rm cyc}(a)\geq\Delta_{D_n}(a).
  \label{eq:parent-cyclic-fixed-comparison}
 \end{equation}
\end{enumerate}
Thus a lower bound proved in the $D_n$-fixed space is also a lower-bound
certificate for the cyclic-envelope gap of this parent protocol.
\end{proposition}

Physically, the protocol explores no more than the $D_n$-fixed sector, and
restricting further to its common cyclic envelope cannot introduce an excitation
below the fixed-sector gap when the unique ground state is shared.  The full
argument is in Appendix~\ref{app:parent-accessibility}.

The number of configurations at energy $u$ is
\begin{equation}
 N_u=\frac{n}{k-u}\binom{k-1}{u}^2,
 \qquad 0\leq u<k.
 \label{eq:cycle-shell-count}
\end{equation}
At $a_f=7$, this shell count implies
\begin{equation}
 1-\langle\psi_{a_f}|\Pi_{U=0}|\psi_{a_f}\rangle
 =O(n^{-5}).
 \label{eq:cycle-endpoint-concentration}
\end{equation}
Here $\Pi_{U=0}$ is the orthogonal projector onto the two-dimensional span
of the configurations satisfying $U(S)=0$.
Appendix~\ref{app:parent-concentration} gives the run decomposition and
ratio bound.

\subsection{A uniform cyclic-envelope gap theorem}

We first separate the cycle geometry from the zero-range kinetics.

\begin{proposition}[Zero-range reduction]
\label{prop:cycle-zrp-reduction}
For every even $n=2k\geq6$ and every $a\geq0$, the gap of $H_a$ in the
$D_n$-fixed space obeys
\begin{equation}
 \Delta_{D_n}(a)
 \geq \frac{2}{n}[1-\cos(2\pi/k)]
 \operatorname{gap}\operatorname{ZRP}(K_k,r_a,k),
 \label{eq:cycle-zrp-reduction}
\end{equation}
where $r_a(0)=0$, $r_a(1)=n^{-a}$, $r_a(j)=1$ for $j\geq2$, and $K_k$
is the uniform mean-field kernel $K_k(i,j)=1/k$ on $k$ boxes.  The notation
$\operatorname{ZRP}(K_k,r_a,k)$ denotes the zero-range process with this
jump kernel, rate function $r_a$, and $k$ particles; its gap is the
Poincar\'e constant of the corresponding generator.  Consequently, any
uniform polynomial lower bound for this mean-field gap gives a uniform
polynomial cyclic-envelope gap certificate for the parent path.
\end{proposition}

The factors in Eq.~\eqref{eq:cycle-zrp-reduction} have distinct physical
origins: $2/n$ is the update-rate normalization, $1-\cos(2\pi/k)$ is the
long-wavelength transport cost of the cycle, and the zero-range gap measures
the remaining occupation-redistribution bottleneck.  The exact isometry,
rate comparison, and normalization are proved in
Appendix~\ref{app:normalized-gap-proof}.

\begin{proposition}[Uniform cyclic-envelope gap]
\label{prop:cycle-uniform-gap}
For every even $n=2k\geq6$ and every $a\geq0$,
\begin{align}
 \operatorname{gap}\operatorname{ZRP}(K_k,r_a,k)
 &\geq 2n^{-a}k^{-3},
 \label{eq:mean-field-bound}\\
 \Delta_{D_n}(a)&\geq1024\,n^{-(a+6)}.
 \label{eq:uniform-cyclic-gap}
\end{align}
Consequently, along $0\leq a\leq7$,
$\min_a\Delta_{D_n}(a)\geq1024\,n^{-13}$.
\end{proposition}

Physically, the polynomial bound accumulates three conservative costs:
normalized subset updates, long-wavelength transport around the cycle, and
redistribution of vacancies between effective boxes.  Its exponent is a
certificate, not a prediction of the finite-size gap scaling.  The complete
comparison is in Appendix~\ref{app:normalized-gap-proof}.

The gap theorem can be converted into an adiabatic statement in the
abstract Hamiltonian-evolution model.  This does not by itself supply a
gate-level implementation of the path.

\begin{corollary}[Conditional adiabatic preparation]
\label{cor:cycle-adiabatic-preparation}
In the abstract Hamiltonian-access model, let $a(s)=7s$, $0\leq s\leq1$,
and evolve from $\ket{J_{n,k}}$ under the restricted path
$H_{a(s)}|_{\mathcal K_J^{\rm par}}$ for total time $T$.  There is a
universal constant $C>0$ such that, for every $0<\epsilon<1$, the sufficient
condition
\begin{equation}
 T\geq C\epsilon^{-1}n^{41}(\log n)^2
 \label{eq:cycle-adiabatic-runtime}
\end{equation}
produces a final state at distance $O(\epsilon)$ from
$\ket{\psi_7}$, up to phase.  Its cover probability is therefore
$1-O(n^{-5})-O(\epsilon)$.  This runtime excludes the costs of preparing the
Dicke state and realizing or simulating $H_{a(s)}$.
\end{corollary}

Physically, only the isolation gap inside $\mathcal K_J^{\rm par}$ enters
the adiabatic theorem; spectral branches in orthogonal symmetry sectors do
not set the preparation time of this protocol.  The large runtime exponent
instead reflects deliberately conservative derivative and inverse-gap
bounds and does not include Hamiltonian-simulation or state-preparation
costs.  The complete adiabatic and measurement estimates are in
Appendix~\ref{app:parent-adiabatic}.

Exact $D_n$-orbit quotients for $n=6,8,\ldots,18$ show finite-size gaps much
larger than the conservative theorem, but are not used to infer scaling.  At
$n=6$ and $a=7$, for example, the certificate is
$1024\,6^{-13}=7.84\times10^{-8}$ whereas the reconstructed gap is
$0.166670$, a factor $2.1\times10^6$ larger.  The analytical bottleneck is
the uniform mean-field estimate $2n^{-a}k^{-3}$; the additional $2/n$
update normalization and cycle transport factor are displayed separately in
Eq.~\eqref{eq:cycle-zrp-reduction}.  The grid and all intermediate values are reported in
Appendix~\ref{app:numerical-protocol} and the archived data.

The two constructions share only the instance family, the cover-measurement
objective, and the orbital language.  No dynamical interpolation or
complexity transfer between them is assumed.

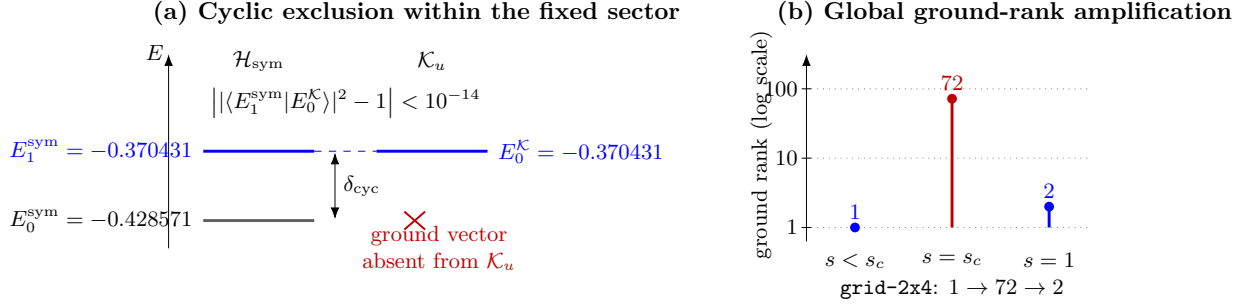
\begin{figure*}[t]
\centering
\resizebox{0.97\textwidth}{!}{%
\begin{tikzpicture}[font=\small,>=Latex]
  \begin{scope}
    \node[anchor=west,font=\bfseries] at (0,3.65) {(a) Cyclic exclusion within the fixed sector};
    \draw[->] (0.35,0.25) -- (0.35,3.05);
    \node at (0.15,3.08) {$E$};
    \node at (1.65,2.95) {$\mathcal H_{\rm sym}$};
    \node at (4.15,2.95) {$\mathcal K_u$};
    \draw[gray!75!black,very thick] (0.85,0.65) -- (2.45,0.65);
    \node[left] at (0.82,0.65) {$E_0^{\rm sym}=-0.428571$};
    \draw[blue,very thick] (0.85,1.65) -- (2.45,1.65);
    \node[left,blue] at (0.82,1.65) {$E_1^{\rm sym}=-0.370431$};
    \draw[blue,very thick] (3.35,1.65) -- (4.95,1.65);
    \node[right,blue] at (4.98,1.65) {$E_0^{\mathcal K}=-0.370431$};
    \draw[blue,dashed] (2.45,1.65) -- (3.35,1.65);
    \draw[red!75!black,thick] (3.75,0.52) -- (4.05,0.78);
    \draw[red!75!black,thick] (4.05,0.52) -- (3.75,0.78);
    \node[red!75!black,align=center] at (4.25,0.25) {ground vector\\absent from $\mathcal K_u$};
    \draw[<->] (2.75,0.67) -- (2.75,1.63) node[midway,right] {$\delta_{\rm cyc}$};
    \node[align=center] at (2.9,2.35) {$\bigl||\langle E_1^{\rm sym}|E_0^{\mathcal K}\rangle|^2-1\bigr|<10^{-14}$};
  \end{scope}
  \begin{scope}[xshift=9.0cm]
    \node[anchor=west,font=\bfseries] at (0,3.65) {(b) Global ground-rank amplification};
    \draw[->] (0.55,0.35) -- (0.55,3.05);
    \node[rotate=90,align=center] at (-0.10,1.75) {ground rank (log scale)};
    \draw (0.48,0.55) -- (0.62,0.55) node[left=2pt] {$1$};
    \draw (0.48,1.55) -- (0.62,1.55) node[left=2pt] {$10$};
    \draw (0.48,2.55) -- (0.62,2.55) node[left=2pt] {$100$};
    \draw[dotted,gray] (0.62,0.55) -- (4.65,0.55);
    \draw[dotted,gray] (0.62,1.55) -- (4.65,1.55);
    \draw[dotted,gray] (0.62,2.55) -- (4.65,2.55);
    \fill[blue] (1.25,0.55) circle (2pt);
    \node[above,blue] at (1.25,0.55) {$1$};
    \node[below,align=center] at (1.25,0.30) {$s<s_c$};
    \draw[very thick,red!75!black] (2.65,0.55) -- (2.65,2.407);
    \fill[red!75!black] (2.65,2.407) circle (2pt);
    \node[above,red!75!black] at (2.65,2.407) {$72$};
    \node[below,align=center] at (2.65,0.30) {$s=s_c$};
    \draw[very thick,blue] (4.05,0.55) -- (4.05,0.851);
    \fill[blue] (4.05,0.851) circle (2pt);
    \node[above,blue] at (4.05,0.851) {$2$};
    \node[below,align=center] at (4.05,0.30) {$s=1$};
    \node[align=center] at (2.65,-0.35) {\texttt{grid-2x4}: $1\to72\to2$};
  \end{scope}
\end{tikzpicture}
}
\caption{Two finite diagnostics of cyclic restriction. (a) At the problem endpoint of \texttt{rnd-b8-t7-p50-s002}, the symmetric ground vector has negligible projection onto $\mathcal K_u$, while the cyclic-envelope ground state coincides, within residual-based precision, with the first symmetric excitation. (b) For \texttt{grid-2x4}, irreducible stoquasticity gives global ground rank one for $s<s_c$, hopping cancellation raises it to $3!\times12=72$ at $s_c=7/15$, and the endpoint rank is two. Panel (b) uses a logarithmic rank axis.}
\label{fig:finite-cyclic-diagnostics}
\end{figure*}

\section{Finite-instance cyclic audit}
\label{sec:numerical-reconstruction}

Figure~\ref{fig:finite-cyclic-diagnostics} isolates the two strongest finite
diagnostics.  They show separately that cyclic exclusion can occur inside
the trivial symmetry sector and that the cancellation point can amplify the
global ground rank without changing the rank followed by the protocol.
All finite-instance Hamiltonians reported here and in
Appendix~\ref{app:numerical-protocol} use the penalty energies $\lambda=5$
and $\mu=10$ in
Eq.~\eqref{eq:hprob-rewrite}.  Every one of the eleven frozen rows uses
$k=k^\star$, verified from the hashed instance registry; hence the
fixed-cardinality success probability of
Eq.~\eqref{eq:fixed-k-cover-probability} is legitimately denoted
$p_{\rm opt}$ below.

At the problem endpoint $H_{\rm prob}$, write $\Delta_{\rm full}$,
$\Delta_{\rm sym}$, and $\Delta_{\rm cyc}$ for the excitation gaps on
$\mathcal H_{\rm addr}$, $\mathcal H_{\rm sym}$, and the common cyclic
envelope $\mathcal K_u$.
The energy costs $\delta_{\rm sym}$ and $\delta_{\rm cyc}$ are those in
Eq.~\eqref{eq:accessibility-cost-rewrite}.  If $\mathcal G_{\rm cyc}$ is
the cyclic ground eigenspace, define its worst-case cover probability by
\begin{equation}
 p_{\rm opt}^{\min}
 =\min_{\substack{\psi\in\mathcal G_{\rm cyc}\\\|\psi\|=1}}
 \langle\psi|P|\psi\rangle,
\end{equation}
where $P$ is the feasible-cover projector from
Section~\ref{sec:rewrite-localization}.

For \texttt{rnd-b8-t7-p50-s002}, the endpoint calculation gives
\begin{equation}
 \begin{aligned}
 |E_0^{\mathcal K}-E_1^{\rm sym}|&<10^{-14},\\
 \left||\langle E_1^{\rm sym}|E_0^{\mathcal K}\rangle|^2-1\right|
 &<10^{-14}.
 \end{aligned}
 \label{eq:cyclic-ground-first-symmetric-excitation}
\end{equation}
The symmetric ground-vector weight in $\mathcal K_u$ is below
$10^{-26}$.  Because $G_B$ is trivial here, both symmetric levels
carry the trivial $S_3$ irrep $[3]$: the exclusion is caused by cyclic
closure, not by a further symmetry sector.  The 19 missing directions may
be chosen shell-local by Eq.~\eqref{eq:shell-complement-decomposition};
their deficits at $\Phi=0,5,15,20,25$ are respectively $2,7,4,4,2$.

For \texttt{grid-2x4}, the global ground rank is one for $s<s_c$, becomes
$3!\times12=72$ at $s_c=7/15$, and is two at the problem endpoint.  This
$1\to72\to2$ comparison collects the rank information that otherwise sits
in separate path and endpoint tables.

Across all eleven frozen instances, five have
$\mathcal K_u\subsetneq\mathcal H_{\rm sym}$, but only one has
$\delta_{\rm cyc}(1)>0$; eight have $\delta_{\rm sym}(1)>0$ and nine have
$\delta_{\rm acc}(1)>0$.  On the audited path grid,
$|\delta_{\rm cyc}(s)|\leq2\times10^{-14}$ for every $s\leq s_c$, as
predicted by Theorem~\ref{thm:shell-cyclic-closure}.  The only positive
endpoint cyclic cost occurs above $s_c$, where the hopping coefficient has
changed sign and the Perron--Frobenius argument no longer applies.  Every
cyclic endpoint ground projector has rank one, with
$p_{\rm opt}^{\min}\geq0.9944$.  These are finite diagnostics, not scaling
evidence.  Appendix~\ref{app:numerical-protocol} gives the full tables,
shell-closure
conditioning, residual-based degeneracy rule, irrep classification, and
OR-Library provenance.

\section{Discussion and conclusion}
\label{sec:rewrite-conclusion}

For a fixed initial state and a symmetry-preserving interpolation, neither
the full spectrum nor, in general, the entire group-fixed spectrum is the
right starting point.  The Hamiltonian may first be restricted to the common
cyclic envelope $\mathcal K_u$, and an adiabatic claim must then identify the
isolated band actually tracked inside it.  Thus
$\operatorname{span}\mathcal R_u[s(\cdot)]\subseteq\mathcal K_u
\subseteq\mathcal H_{\rm sym}$ is the central hierarchy: changing the
initial state, generator algebra, schedule, or preserved symmetry can change
the relevant spectrum and gap.  States outside $\mathcal K_u$ are
inaccessible to the generator family, whereas inclusion in $\mathcal K_u$
alone is not a state-preparability theorem.

Within this framework, orbit invariance identifies unlabelled candidate
covers as the natural sites of the fixed-sector quotient, while the shell
closure determines which amplitude patterns among those sites are excluded
or retained by the endpoint algebra.  The sector-resolved localization
proposition
certifies cover-measurement probability with the required kinetic floor.
Exact orbit quotients separate ground multiplicity, excitation gaps,
isolation gaps, and restriction costs, and expose coincidences between
irreducible sectors that are dark to the symmetric protocol.  At the
hopping-cancellation point, the retained-walk interpolation has an exact global
multiplicity closure for every feasible fixed-cardinality instance.  The
even cycle is a cover-transitive illustration whose fixed excitation gap is
at least $3\lambda/7$ and approaches $\lambda/2$ from below at that point.
A distinct Johnson--Metropolis parent
path has a uniformly polynomial cyclic-envelope gap bounded below by
$1024\,n^{-13}$ along the path, and consequently a
conditional polynomial adiabatic runtime in the abstract
Hamiltonian-access model.

These results establish a structural separation, not a practical quantum
speedup.  The cycle instances are classically solvable, Dicke-state
preparation has a cost that must be counted, and the Gibbs parent endpoint is
not the retained-walk problem Hamiltonian $H_{\rm prob}$.  Likewise, the address count
$k\lceil\log_2n_B\rceil$ is not a complete hardware resource estimate:
incidence access, flags, ancillas, and gate depth require a separate
implementation analysis.

A natural next step is a decorated family whose classical cover structure
is nontrivial while its joint orbit quotient, dark global branch, and
polynomial tracked isolation gap remain stable under controlled
symmetry-preserving perturbations.  More broadly, spectral claims for
symmetric Hamiltonian algorithms should identify not only a Hamiltonian and
its global gap, but also the initial state, preserved symmetry, common cyclic
envelope, explicit schedule, and isolated band actually followed by the
protocol.

\section*{Acknowledgements}
This study was financed, in part, by the São Paulo Research Foundation
(FAPESP), Brazil, grant 2026/04387-7.  The author thanks Marcos César de
Oliveira for supervision and discussions throughout this project.

\section*{Author contributions}
Fabrício de Souza Luiz conceived the project, developed the mathematical
analysis, implemented and validated the numerical calculations, and wrote
the manuscript.  Large-language-model tools from OpenAI were used to assist
with editorial reorganization, language revision, bibliographic discovery,
LaTeX editing, and the development and checking of reproducibility code.
The author reviewed all model-assisted material, verified the mathematical
and numerical results, and takes full responsibility for the content of the
work.

\section*{Code and data availability}
The frozen instance registry, source code, residual-based numerical checks,
and machine-readable tables accompanying this preprint are available in the
versioned repository
\url{https://github.com/fsluiz/manuscript-encoding-symmetry-limits}.
The version-specific archival record is available at
\url{https://doi.org/10.5281/zenodo.22142799}.
Original software is released under the MIT License; original data and
documentation are released under CC BY 4.0.  Third-party materials, including
OR-Library incidence data and the journal class file, retain their source
licenses and provenance.

\bibliographystyle{quantum}
\bibliography{refs}

\appendix
\section{Proof of sector localization}
\label{app:sector-localization}

We prove Proposition~\ref{prop:sector-localization-rewrite} and record why
its sector hypothesis cannot be silently replaced by cyclic restriction.
The variational principle applied inside $\Ran\Pi_\gamma$ gives
$E_0^\gamma\leq\eta_\gamma$.  Combining this with
Eq.~\eqref{eq:q-lower-rewrite} yields
\begin{equation}
 \Gamma_\gamma\geq m-\kappa-\eta_\gamma,
\end{equation}
which is Eq.~\eqref{eq:sector-separation-rewrite}.

For a normalized sector ground vector, write
$p=P_\gamma\psi_\gamma$ and $q=Q_\gamma\psi_\gamma$.  Since $P$, $Q$,
and $\Pi_\gamma$ commute, these are orthogonal components and the lower
block of the eigenvalue equation is
\begin{equation}
 (Q_\gamma H_{\rm prob}Q_\gamma-E_0^\gamma)q=-B_\gamma p.
 \label{eq:app-sector-block}
\end{equation}
If $Q_\gamma=0$, then $q=0$ identically.  Otherwise positivity of
$\Gamma_\gamma$ makes the compressed operator on the left invertible with
inverse norm at most $\Gamma_\gamma^{-1}$.  Therefore
\begin{equation}
 \|q\|\leq\frac{\|B_\gamma\|}{\Gamma_\gamma}\|p\|.
\end{equation}
Using $\|p\|^2+\|q\|^2=1$ and solving for $\|q\|^2$ gives
Eq.~\eqref{eq:sector-bound-rewrite}.  Replacing $\Gamma_\gamma$ by the
positive lower bound $m-\kappa-\eta_\gamma$ weakens the right-hand side and
proves the stated certificate.

For finite penalties, the proposition controls rather than eliminates
leakage; exact support on covers requires an infinite-penalty limit or a
special decoupling.  Moreover, if $\Pi_{\mathcal K}$ denotes the orthogonal
projector onto the common cyclic envelope $\mathcal K_u$, it need not commute
with $P$, so
$P\Pi_{\mathcal K}$ need not be an orthogonal
projector.  The proposition therefore applies directly to genuine
commuting symmetry sectors.  Cyclic-envelope solution probability must be
evaluated directly or bounded with a separately constructed commuting
projector.

\section{Joint representation and orbit quotient}
\label{app:joint-representation}

The incidence automorphism group $G_A$ acts on the address Hilbert space
only through its base projection.  If
$\rho(\pi_B,\pi_T)=\pi_B$, then elements with $\pi_B=\mathrm{id}$ form the
kernel, and the faithfully represented group is
$G_B\simeq G_A/\ker\rho$.  Since the $S_k$ register action commutes with the
$G_B$ base action, complete reducibility gives
\begin{equation}
 \mathcal H_{\rm addr}\simeq
 \bigoplus_{\alpha,\beta}
 V_\alpha^{S_k}\otimes V_\beta^{G_B}\otimes
 \mathcal M_{\alpha\beta}.
 \label{eq:app-joint-decomposition}
\end{equation}
Every invariant Hamiltonian lies in the commutant and hence has the form in
Eq.~\eqref{eq:commutant-rewrite}.  The identity factors generate Schur
multiplicities; equality of eigenvalues belonging to different
$(\alpha,\beta)$ blocks is not enforced by this statement.

\subsection{Orbit invariance and enumeration}

Register permutations leave the multiset
$\{\!\{b_1,\ldots,b_k\}\!\}$ unchanged, so they preserve both $U$ and
$D$.  A base automorphism $g\in G_B$ is a bijection; hence
$b_r=b_s$ if and only if $g(b_r)=g(b_s)$, proving invariance of $D$.
For $U$, choose any task permutation $\pi_T$ such that
$(g,\pi_T)\in G_A$.  If $S$ is the underlying set of selected bases, write
\begin{equation}
 \mathcal U(S)=\{t\in\mathcal T:A_{bt}=0\ \text{for every }b\in S\}.
\end{equation}
The incidence-automorphism identity gives
$\mathcal U(gS)=\pi_T\mathcal U(S)$, so the two uncovered-task sets have
the same cardinality.  This proves Lemma~\ref{lem:orbit-invariance}.  The
cardinality is independent of which compatible $\pi_T$ is chosen, as it
must be because only the faithfully represented base action enters the
Hilbert space.

For completeness, let $\mathcal O$ and $\mathcal O'$ be computational-basis
orbits of the joint action and choose $x\in\mathcal O$.  Invariance gives
\begin{align}
 \langle\mathcal O'|H|\mathcal O\rangle
 &=\frac{1}{\sqrt{|\mathcal O||\mathcal O'|}}
 \sum_{x'\in\mathcal O}\sum_{y\in\mathcal O'}H_{yx'}\\
 &=\sqrt{\frac{|\mathcal O|}{|\mathcal O'|}}
 \sum_{y\in\mathcal O'}H_{yx},
\end{align}
which proves Eq.~\eqref{eq:orbit-quotient-rewrite}.  The orbit states are
therefore an exact orthonormal basis of the fixed space, not an effective
approximation.

After quotienting by register permutations, basis vectors are multisets of
size $k$ over the bases.  If $g\in G_B$ has $c_\ell(g)$ cycles
of length $\ell$, the trace of $g$ on
the symmetric tensor power $\operatorname{Sym}^k(\mathbb C^{n_B})$ is the
coefficient
\begin{equation}
 [z^k]\prod_{\ell\geq1}(1-z^\ell)^{-c_\ell(g)}.
\end{equation}
The Cauchy--Frobenius orbit-counting lemma, equivalently the
P\'olya--Redfield cycle-index evaluation for multisets, averages this
character over $G_B$ and projects onto the trivial irrep.  This gives
Eq.~\eqref{eq:polya-rewrite}.  It counts
$\mathcal H_{\rm sym}$ only.  Whether $\ket u$ is cyclic for the restricted
endpoint algebra is a separate closure calculation.

\subsection{Proof of the shell-resolved cyclic closure}

The endpoint operators satisfy
\begin{equation}
 H_{\rm init}=kI-\frac{M}{n_B},
 \qquad
 H_{\rm walk}=\frac{M-kI}{n_B-1}.
 \label{eq:app-two-generator-identities}
\end{equation}
Thus $M=n_B(kI-H_{\rm init})$ belongs to the endpoint algebra,
$H_{\rm walk}$ follows from $M$, and
$V=H_{\rm prob}-H_{\rm walk}$ also belongs to it.  Conversely,
Eq.~\eqref{eq:app-two-generator-identities} reconstructs both endpoints
from $M,V,I$.  This proves Lemma~\ref{lem:two-generator-reduction}.

Because $\widehat V=\sum_j\varphi_jR_j$, invariance under $\widehat V$ is
equivalent to invariance under every $R_j$: each $R_j$ is a Lagrange
interpolation polynomial in $\widehat V$.  The initial shell vectors belong
to the cyclic envelope since
\begin{equation}
 \widehat V^i\boldsymbol w
 =\sum_{j=1}^p\varphi_j^iR_j\boldsymbol w,
 \qquad 0\leq i<p,
\end{equation}
and the Vandermonde matrix of the distinct $\varphi_j$ is invertible.
Every $R_j\boldsymbol w$ is nonzero by
Eq.~\eqref{eq:uniform-orbit-coordinates}.
The iteration in Eq.~\eqref{eq:shell-cyclic-iteration} therefore begins
inside $\mathcal K_u$, remains there, and stabilises after finitely many
steps.  Its limit contains $\boldsymbol w$, is invariant under every $R_j$ and under
$\widehat M$, hence under $\widehat V$ and the complete endpoint algebra.
Minimality of the cyclic envelope proves equality.  The $p$ nonzero shell
vectors are mutually orthogonal, which proves
Eq.~\eqref{eq:shell-lower-bound}.  If every shell is a singleton, those
vectors are nonzero multiples of all orbit-basis vectors, proving the
injectivity criterion and its contrapositive.

Since every $R_j$ is an orthogonal projector in the endpoint algebra,
$R_j\mathcal K_u\subseteq\mathcal K_u$.  The mutually orthogonal projectors
sum to the identity on the orbit space, so
$\mathcal K_u=\bigoplus_jR_j\mathcal K_u$.  Taking the orthogonal complement
inside $\bigoplus_j\operatorname{Ran}R_j$ gives the second identity in
Eq.~\eqref{eq:shell-complement-decomposition}.  Thus a basis of inaccessible
directions may always be chosen shell-local, although this statement alone
does not assign an orientation or current interpretation to those
directions.

It remains to prove Eq.~\eqref{eq:zero-cyclic-cost-before-cancellation}.
For $s$ below the unique zero of $c(s)$, every off-diagonal entry of the
computational-basis Hamiltonian is nonpositive.  Equation
\eqref{eq:orbit-quotient-rewrite} preserves that sign, and the orbit hopping
graph is connected: a sequence of one-register substitutions connects any
two size-$k$ multisets.  After a sufficiently large scalar shift,
Perron--Frobenius~\cite{HornJohnson2012} therefore gives a simple symmetric
ground vector $\psi_0(s)$ with strictly positive orbit coordinates.  Its
overlap with $\boldsymbol w$ is nonzero.  The rank-one spectral projector onto
$\psi_0(s)$ is a polynomial in the finite-dimensional quotient Hamiltonian,
which belongs to the algebra generated by $\widehat M$ and $\widehat V$.
Applying that projector to $\boldsymbol w$ proves
$\psi_0(s)\in\mathcal K_u$.

At the zero of $c(s)$, the Hamiltonian is a scalar plus a positive multiple
of $\widehat V$.  If $R_{\min}$ denotes its lowest nonempty potential shell,
then $R_{\min}w\ne0$ and belongs to $\mathcal K_u$.  It is a ground vector,
so the symmetric and cyclic ground energies again agree.  Continuity also
covers $s=0$.  This completes the proof of
Theorem~\ref{thm:shell-cyclic-closure}.

\subsection{Proof of the cancellation theorem}

Substitution of Eq.~\eqref{eq:app-two-generator-identities} into the linear
path gives Eq.~\eqref{eq:path-transport-decomposition}.  Solving $c(s)=0$
gives $s=s_c$, and direct evaluation gives $a(s_c)=ks_c$ and
$s_cV$ for the remaining potential term.  This proves
Eq.~\eqref{eq:cancellation-identity}.

In the computational basis, $V$ is diagonal with entries $\Phi$; by
Lemma~\ref{lem:orbit-invariance}, each potential shell is a union of joint
orbits.  Therefore every $\varphi_j$ occurs in the full and fixed-space
spectra.  It also occurs in the cyclic restriction because the nonzero
vector $R_j\boldsymbol w$ belongs to $\mathcal K_u$ and is an eigenvector of
$H(s_c)$.  No other eigenvalue is possible in any restriction, proving
Eqs.~\eqref{eq:cancellation-levels} and
\eqref{eq:cancellation-gap}.  If there is no second shell, the excitation
gap is assigned the convention $+\infty$.

For a feasible exact-$k$ subset, all $k!$ orderings are distinct
computational-basis ground vectors.  Conversely, zero potential requires
an exact-$k$ cover with no repeated bases, so this accounts for the complete
full ground rank.  Register quotienting turns those tuples into subsets,
and quotienting by $G_B$ leaves one fixed-space vector per feasible-cover
orbit.  Finally, $R_1w$ supplies at least one cyclic ground vector, while
$\mathcal K_u\subseteq\mathcal H_{\rm sym}$ gives the upper rank bound.
This proves Eqs.~\eqref{eq:cancellation-full-ranks} and
\eqref{eq:cancellation-cyclic-rank}, and hence
Theorem~\ref{thm:orbit-resolved-cancellation}.

\section{Stability and the finite \texorpdfstring{$D_4$}{D4} diagnostic}

\subsection{Proof of the stability theorem}
\label{app:dark-stability}

Simple isolated reduced eigenvalues of a differentiable Hermitian family
define differentiable local branches; the identity factor on an irrep may
replicate such a branch without changing its scalar energy.  Apply the
implicit-function theorem to
$F(s,\varepsilon)=E_a(s,\varepsilon)-E_b(s,\varepsilon)$ and use
Eq.~\eqref{eq:transverse-dark-hypothesis}.  This gives a unique nearby root
$s_\star(\varepsilon)=s_\star+O(|\varepsilon|)$.  Commutation with the
symmetry projectors makes every inter-sector matrix element vanish, so the
equality cannot become an avoided crossing.  Finally, Weyl's bound moves
each eigenvalue of the $a$ block by at most
$|\varepsilon|\|R(s)\|$.  An internal spectral distance can shrink by at
most twice this quantity, which proves the isolation-gap statement in
Theorem~\ref{thm:dark-crossing-stability}.

\subsection{Catalyst and numerical audit}
\label{app:d4-catalyst}

The frozen instance has bases and tasks labelled $0,\ldots,7$ and incidence
matrix
\begin{equation}
 A=\begin{pmatrix}
 1&1&0&0&1&0&0&0\\
 1&1&1&0&0&1&0&0\\
 0&1&1&1&0&0&1&0\\
 0&0&1&1&0&0&0&1\\
 1&0&0&0&1&1&0&0\\
 0&1&0&0&1&1&1&0\\
 0&0&1&0&0&1&1&1\\
 0&0&0&1&0&0&1&1
 \end{pmatrix}.
 \label{eq:d4-incidence-matrix}
\end{equation}
It is the closed-neighbourhood incidence relation of the $2\times4$ vertex
grid.  Because task labels may be permuted independently of base labels,
its effective incidence group has order eight and is isomorphic to $D_4$;
this need not equal the geometric automorphism group obtained by forcing the
same permutation on both colours.  For $k=3$, the $S_3\times D_4$ fixed
quotient and the cyclic closure from $\ket u$ both have dimension $22$,
compared with $512$ in the ordered address space.  The latter equality is an output of the closure calculation,
not an assumption based on symmetry.  The orbit-isometry,
endpoint-invariance, and cyclic-invariance residuals are below $10^{-13}$.

For the catalyst in Eqs.~\eqref{eq:d4-catalyst-operator}--
\eqref{eq:d4-catalyst-path}, set $\chi=2$.  The remaining global crossing
is at $s\simeq0.806454$.  The two branches belong to different $D_4$ irreps,
Eq.~\eqref{eq:dark-coupling-rewrite} has residual
$4.7\times10^{-14}$, the reduced-branch slope difference is approximately
$-4.54$, and the $D_4$-trivial gap there is approximately $0.258$.  The
nontrivial branch is a rank-two $E$ multiplet; the derivative restricted to
that multiplet is scalar to numerical precision.  The trivial-sector gap
was evaluated on a uniform 2001-point grid
and locally minimized in every grid bracket; its minimum is $0.0414$ at
$s\simeq0.874587$.  The crossing between the trivial floor and the
orthogonal-complement floor was bracketed on 401 points and refined by a
scalar root solve.  These are residual-tested finite diagnostics, not
interval-certified bounds, and are not used in the asymptotic theorem.

\section{Parent cyclic envelope and endpoint concentration}

\subsection{Cyclic-envelope inclusion}
\label{app:parent-accessibility}

We prove Proposition~\ref{prop:parent-cyclic-accessibility}.  Because
$\mathfrak A_{\rm par}$ is a unital star-algebra,
$\mathfrak A_{\rm par}\ket{J_{n,k}}$ is invariant under every element of
the algebra and its adjoint, hence is reducing.  Every $H_a$ commutes with
$D_n$, and $\ket{J_{n,k}}$ is $D_n$-fixed, proving the first item.

All rates in Eq.~\eqref{eq:johnson-metropolis} are positive on the connected
Johnson graph.  The zero eigenvalue of $H_a$ is consequently simple, with
ground vector Eq.~\eqref{eq:cycle-parent-ground}.  In finite dimension its
spectral projector $P_a=\ket{\psi_a}\!\bra{\psi_a}$ is a polynomial in
$H_a$ and belongs to $\mathfrak A_{\rm par}$.  Positivity gives
$\langle\psi_a|J_{n,k}\rangle>0$, so
$P_a\ket{J_{n,k}}$ is a nonzero multiple of $\ket{\psi_a}$.  Finally, both
restrictions have this unique ground state, while
$\mathcal K_J^{\rm par}\subseteq\mathcal H^{D_n}$.  Rayleigh--Ritz
minimization over the smaller orthogonal complement proves
Eq.~\eqref{eq:parent-cyclic-fixed-comparison}.

\subsection{Endpoint concentration}
\label{app:parent-concentration}

At energy $u$, occupied and vacant vertices decompose into the same number
$k-u$ of positive cyclic runs.  Choosing the marked start and the two run
compositions gives
\begin{equation}
 N_u=\frac{n}{k-u}\binom{k-1}{u}^2,
 \qquad0\leq u<k,
\end{equation}
which is Eq.~\eqref{eq:cycle-shell-count}.  At $a_f=7$, the $u=1$
contribution is $N_1n^{-7}=n(k-1)n^{-7}=O(n^{-5})$.  Consecutive terms
$N_un^{-7u}$ have ratio at most $2k^2n^{-7}=O(n^{-5})$.  Summing the
resulting geometric tail proves
Eq.~\eqref{eq:cycle-endpoint-concentration}.

\subsection{Adiabatic derivative bounds}
\label{app:parent-adiabatic}

A Johnson move removes one selected cycle vertex and inserts one unselected
vertex, so $|U(T)-U(S)|\leq2$, and each subset has
$k(n-k)=k^2$ neighbours.  Differentiating the matrix elements in
Eqs.~\eqref{eq:johnson-metropolis} and \eqref{eq:cycle-parent}, then bounding
the operator norm by the maximum absolute row sum, gives, for constants
$c_1,c_2>0$ independent of $n$,
\begin{equation}
 \|\partial_sH_{7s}\|\leq c_1n\log n,
 \qquad
 \|\partial_s^2H_{7s}\|\leq c_2n(\log n)^2.
 \label{eq:cycle-parent-derivative-bounds}
\end{equation}
Proposition~\ref{prop:parent-cyclic-accessibility} and
Eq.~\eqref{eq:uniform-cyclic-gap} give a simple ground state and minimum
cyclic-envelope gap at least $1024n^{-13}$.

Apply the quantitative gapped adiabatic bound
\cite{JansenRuskaiSeiler2007} to
$H_{7s}|_{\mathcal K_J^{\rm par}}$; restriction cannot increase either
derivative norm.  The three contributions scale as
$n^{27}\log n$, $n^{27}(\log n)^2$, and
$n^{41}(\log n)^2$, divided by $T$.  The last dominates and proves
Eq.~\eqref{eq:cycle-adiabatic-runtime}.  The final measurement statement
follows from Eq.~\eqref{eq:cycle-endpoint-concentration} and continuity of
measurement probabilities in state distance.

\section{Normalized comparison proof for the parent-path gap}
\label{app:normalized-gap-proof}

This appendix fixes the generator convention used in
Propositions~\ref{prop:cycle-zrp-reduction} and
\ref{prop:cycle-uniform-gap} and records each comparison factor.  For a
reversible generator $\mathcal L$ with stationary law $\nu$, write
\begin{align}
 \mathcal E_{\mathcal L}(f,f)
 &=-\langle f,\mathcal Lf\rangle_\nu,\\
 \operatorname{gap}(-\mathcal L)
 &=\inf_{\operatorname{Var}_\nu(f)>0}
 \frac{\mathcal E_{\mathcal L}(f,f)}
 {\operatorname{Var}_\nu(f)}.
 \label{eq:app-dirichlet-convention}
\end{align}

\begin{lemma}[Three-box heat-bath contraction]
\label{lem:three-box-contraction}
Let $w_0=\tau$, $w_j=1$ for $j\geq1$, with $0<\tau\leq1$.  On weak
compositions of any total $m$ into three boxes, let one heat-bath step keep
one uniformly chosen coordinate and resample the other two from their
conditional product law.  The continuous-time generator
$\mathcal L^*_{3,m}$ obtained by subtracting the identity has gap at least
$1/3$.
\end{lemma}

\subsection{Marked configurations and the cycle zero-range chain}
\label{app:marked-zrp}

Put $\tau=n^{-a}$ and let
$\Omega_{n,k}=\{S\subset\mathbb Z_n:|S|=k\}$.  Let $\mathcal Q_a$ denote
the Markov generator with the rates in
Eq.~\eqref{eq:johnson-metropolis} and stationary law
$\pi_a(S)\propto\tau^{U(S)}$; its symmetric discriminant is $H_a$.  If
$D_{\pi_a}$ is diagonal with entries
$\pi_a(S)$, then
\begin{equation}
 H_a=-D_{\pi_a}^{1/2}\mathcal Q_aD_{\pi_a}^{-1/2}.
 \label{eq:app-discriminant-similarity}
\end{equation}
The unitary map
$f\mapsto\sum_S\sqrt{\pi_a(S)}f(S)\ket S$ intertwines the $D_n$ actions.
Thus the gap of $H_a$ in the fixed space is the Poincar\'e constant of
$\mathcal Q_a$ on $D_n$-invariant observables.

Introduce the marked space
\begin{align}
 \widehat\Omega_{n,k}
 &=\{(S,x):S\in\Omega_{n,k},\ x\in S\},\\
 \widehat\pi_a(S,x)&=\frac{\pi_a(S)}{k}.
\end{align}
Lift every Johnson transition that replaces $y\in S$ by $z\notin S$ to the
marked space by leaving the mark unchanged when $x\ne y$ and replacing the
mark by $z$ when $x=y$.  Thus the mark follows its labelled particle.  This
lifted generator is reversible with respect to $\widehat\pi_a$.  For the
mark-independent lift $\widehat f(S,x)=f(S)$, its variance and Dirichlet
form are exactly those of $f$ under the original chain.

Starting at the marked particle $x$, list the occupied vertices in cyclic
order and let $g_i$ be the number of vacant vertices after particle $i$.
This gives a bijection
\begin{equation}
 (S,x)\longleftrightarrow(x,g),
 \qquad x\in\mathbb Z_n,
 \end{equation}
between marked configurations and anchored weak compositions satisfying
\begin{equation}
 g_i\geq0,\qquad \sum_{i=1}^kg_i=k,
 \qquad U(g)=\#\{i:g_i=0\}.
 \label{eq:app-gap-composition}
\end{equation}
Changing the marked particle cyclically shifts $g$, while rotating or
reflecting the subset changes the anchor and possibly the orientation.
Here is the first indispensable use of dihedral invariance: only for a
$D_n$-invariant observable $f(S)$ is its marked lift independent of the
anchor and representable by a cyclic-and-reflection-invariant function
$F(g)$.  When the marked particle itself moves, the anchor $x$ changes by
the rule above, but $F$ is unaffected by that change.  Without $D_n$
invariance the lift would be a function of $(x,g)$ and the
composition-only comparison below would not follow.

Define
\begin{equation}
 r_a(0)=0,\qquad r_a(1)=\tau,\qquad r_a(j)=1\quad(j\geq2).
 \label{eq:app-zrp-rates}
\end{equation}
The product weight of a composition is
\begin{equation}
 \prod_{i=1}^k\prod_{j=1}^{g_i}r_a(j)^{-1}
 =\tau^{-\#\{i:g_i>0\}}
 =\tau^{-k}\tau^{U(g)}.
 \label{eq:app-zrp-weight}
\end{equation}
Let $\nu_a$ be the probability law on weak compositions obtained by
normalizing this product weight.  The irrelevant factor $\tau^{-k}$ shows
that the pushforward of the marked
law after dropping the anchor is exactly the stationary composition law:
every unanchored composition $g$ has precisely $n$ anchored preimages, all
with the same weight.  In particular,
\begin{equation}
 \operatorname{Var}_{\pi_a}(f)
 =\operatorname{Var}_{\widehat\pi_a}(f)
 =\operatorname{Var}_{\nu_a}(F).
 \label{eq:app-variance-isometry}
\end{equation}

Let $P_{C_k}(i,j)=1/2$ when $j=i\pm1\pmod k$ and zero otherwise, and let
$e_i$ be the $i$th coordinate vector.  We use the Hermon--Salez convention
\begin{equation}
 (\mathcal L_{P,r}F)(g)
 =\sum_{i,j}r(g_i)P(i,j)
 [F(g-e_i+e_j)-F(g)].
 \label{eq:app-zrp-generator}
\end{equation}
The notation $\operatorname{ZRP}(P,r,k)$ below denotes the zero-range generator
$\mathcal L_{P,r}$ restricted to configurations with $k$ particles.
Multiplication by $2/n$ makes every oriented cycle transfer occur at rate
$r_a(g_i)/n$.  Such a transfer is an adjacent Johnson move.  If $g_i=1$
and the receiving box is positive, it creates one new zero and has the
Metropolis rate $\tau/n$; if the receiving box is zero, its Metropolis rate
is $1/n\geq\tau/n$.  For $g_i\geq2$, the move does not increase $U$ and
both rates equal $1/n$.  Hence the full Johnson Dirichlet form dominates
the local form and, under the lift above,
\begin{equation}
 \mathcal E_{\mathcal Q_a}(f,f)
 \geq\frac2n
 \mathcal E_{\operatorname{ZRP}(P_{C_k},r_a,k)}(F,F).
 \label{eq:app-local-domination}
\end{equation}
The equality of variances is exact; the inequality comes only from dropping
nonlocal Johnson edges and, in one local case, lowering a Metropolis rate.

Let $K_k(i,j)=1/k$.  Corollary~3 of Hermon and Salez
\cite{HermonSalez2019}, in their site-independent-rate convention, states
the pointwise Dirichlet-form comparison
\begin{multline}
 \mathcal E_{\operatorname{ZRP}(P_{C_k},r_a,k)}(F,F)\\
 \geq[1-\cos(2\pi/k)]
 \mathcal E_{\operatorname{ZRP}(K_k,r_a,k)}(F,F),
 \label{eq:app-hermon-salez}
\end{multline}
because $1-\cos(2\pi/k)$ is the Poincar\'e constant of $P_{C_k}$ and that
of $K_k$ is one.  Their comparison applies to every observable before any
symmetry restriction and does not require monotone rates.  In the present
case the rates are nevertheless nondecreasing,
$0=r_a(0)\leq r_a(1)=\tau\leq r_a(j)=1$ for $j\geq2$.  The dihedral
restriction was already needed to define $F(g)$; restricting the subsequent
variational infimum can only increase it.  Equations
\eqref{eq:app-variance-isometry}--\eqref{eq:app-hermon-salez} prove
Eq.~\eqref{eq:cycle-zrp-reduction} with its factor $2/n$.

\subsection{Three-box contraction}
\label{app:three-box-proof}

Set $w_0=\tau$ and $w_j=1$ for $j\geq1$.  For $0<z<1$, the probability
sequence $p_z(j)\propto w_jz^j$ is log-concave: the only nontrivial
inequality is $w_1^2\geq w_0w_2$, which is precisely $\tau\leq1$.
Conditioning two independent variables with this law to have sum $s$
removes the factor $z^s$ and gives
\begin{equation}
 \nu_s(u)=\frac{w_u w_{s-u}}
 {\sum_{v=0}^s w_v w_{s-v}},
 \qquad0\leq u\leq s.
 \label{eq:app-split-law}
\end{equation}
The discrete Efron monotonicity theorem~\cite{Efron1965} implies that
$U_s\sim\nu_s$ is stochastically nondecreasing in $s$.  Write
$\preceq_{\rm st}$ for stochastic domination.  Symmetry under
$u\mapsto s-u$ gives, for $t=s+d$,
\begin{equation}
 U_s\preceq_{\rm st}U_t\preceq_{\rm st}U_s+d.
 \label{eq:app-bounded-stochastic-order}
\end{equation}
The common-quantile coupling therefore obeys
$0\leq U_t-U_s\leq d$ almost surely.  The two increments in the coupled
splits are $U_t-U_s$ and $d-(U_t-U_s)$, so both are nonnegative.

On the three-box state space use
$d_1(x,y)=\frac12\sum_i|x_i-y_i|$.  Couple two heat-bath steps by holding
the same uniformly selected coordinate $i$.  The two resampled pair totals
differ by $d=|x_i-y_i|$, and the coupling above gives pairwise
$\ell^1$ distance $d$.  Together with the held-coordinate distance, the
post-update $d_1$ distance is at most $|x_i-y_i|$.  Let $W_{1,d_1}$ denote
the $1$-Wasserstein distance induced by $d_1$.  Thus the discrete heat-bath
kernel $\mathsf P$ satisfies
\begin{equation}
 W_{1,d_1}(\mathsf P(x,\cdot),\mathsf P(y,\cdot))
 \leq\frac13\sum_i|x_i-y_i|=\frac23d_1(x,y).
 \label{eq:app-three-box-wasserstein}
\end{equation}
For a function $f$, let $\operatorname{Lip}(f)
=\sup_{x\ne y}|f(x)-f(y)|/d_1(x,y)$ be its Lipschitz seminorm.  It follows
that $\operatorname{Lip}(\mathsf Pf)\leq(2/3)\operatorname{Lip}(f)$.  Every
nonconstant eigenfunction on this finite metric space has positive
Lipschitz seminorm, so all nonconstant eigenvalues of the reversible kernel
have absolute value at most $2/3$.  Therefore
$\mathcal L^*_{3,m}=\mathsf P-I$ has gap at least $1/3$, uniformly in the
conserved total $m$.  This proves
Lemma~\ref{lem:three-box-contraction}.

\subsection{Caputo recursion with the present normalization}
\label{app:heat-bath-recursion}

For $N$ boxes and conserved total $m$, let
\begin{align}
 \Omega_{N,m}
 &=\{g\in\mathbb Z_{\geq0}^{N}:\textstyle\sum_i g_i=m\},\\
 \nu_{N,m}(g)&=Z_{N,m}^{-1}\prod_{i=1}^{N}w_{g_i}.
\end{align}
Here $Z_{N,m}$ is the normalizing partition function.
Let $E_{ij}$ denote conditional expectation under $\nu_{N,m}$ given all
coordinates outside the unordered pair $\{i,j\}$.  The complete-graph
pair heat-bath generator is normalized as
\begin{equation}
 \mathcal L_{\rm HB}^{N,m}
 =\frac1N\sum_{1\leq i<j\leq N}(E_{ij}-I).
 \label{eq:app-heat-bath-generator}
\end{equation}
For $N=3$, the discrete kernel $I+\mathcal L_{\rm HB}^{3,m}$ keeps one
uniformly chosen coordinate and resamples the remaining pair.  It is exactly
the kernel of Lemma~\ref{lem:three-box-contraction}, not a rescaled version.

Choose any $0<z<1$ and let the one-box law be
$\mu_z(j)\propto w_jz^j$.  Then $\nu_{N,m}$ is exactly the product law
$\mu_z^{\otimes N}$ conditioned on $\sum_i g_i=m$, because the factor
$z^m$ cancels.  Consequently, after conditioning on all coordinates outside
a set $A$, the coordinates in $A$ again have their product law conditioned
only on the remaining total $m-\sum_{i\notin A}g_i$.  This is precisely the
``non-interference property'' defined before Eq.~(2.2) of
Caputo~\cite{Caputo2008}.

For $m=0$ the conditioned law is
a point mass, contributes no conditional variance, and its gap is assigned
the convention $+\infty$.  With
\begin{equation}
 \underline\lambda_3
 =\inf_{m\geq1}\operatorname{gap}
 (-\mathcal L_{\rm HB}^{3,m}),
\end{equation}
Theorem~3.1 of Caputo~\cite{Caputo2008}, in the normalization
\eqref{eq:app-heat-bath-generator}, gives
\begin{equation}
 \operatorname{gap}(-\mathcal L_{\rm HB}^{N,m})
 \geq(3\underline\lambda_3-1)
 \left(1-\frac2N\right)+\frac1N.
 \label{eq:app-caputo-recursion}
\end{equation}
Lemma~\ref{lem:three-box-contraction} gives
$\underline\lambda_3\geq1/3$.  The first term on the right of
Eq.~\eqref{eq:app-caputo-recursion} is therefore nonnegative, and
\begin{equation}
 \operatorname{gap}(-\mathcal L_{\rm HB}^{N,m})\geq\frac1N
 \label{eq:app-heat-bath-gap}
\end{equation}
for every conserved total.  In the application below, $N=m=k$.

\subsection{Conditional two-box comparison}
\label{app:mean-field-comparison}

For the mean-field kernel $K_k(i,j)=1/k$, decompose its zero-range generator
into unordered pairs:
\begin{equation}
 \mathcal L_{\operatorname{ZRP}(K_k,r_a,m)}
 =\frac2k\sum_{1\leq i<j\leq k}\mathcal L_{ij}^{(2)}.
 \label{eq:app-mean-field-pair-decomposition}
\end{equation}
Here $\mathcal L_{ij}^{(2)}$ is the conditional two-box zero-range generator
with kernel $K_2$, whose two rows are $(1/2,1/2)$.  Thus a unit leaves either
member of the pair toward the other at rate $r_a/2$.  The prefactor $2/k$
in Eq.~\eqref{eq:app-mean-field-pair-decomposition} converts this to the
mean-field rate $r_a/k$.

Condition on pair total $s$.  For $s\geq2$, the stationary weights on
$\{0,\ldots,s\}$ are $\tau$ at the two endpoints and one in the interior,
with
\begin{equation}
 Z_s=s-1+2\tau.
\end{equation}
Every nearest-neighbour edge conductance of
$\mathcal L_{ij}^{(2)}$ is at least $\tau/(2Z_s)$.  For the canonical path
between $u<v$, Cauchy--Schwarz gives a factor $v-u\leq s$.  For every edge,
the stationary masses on its two sides have product at most $1/4$.
Let $\mathcal E_s$ be the Dirichlet form of this conditional two-box
generator.  Consequently
\begin{equation}
 \operatorname{Var}_{\nu_s}(f)
 \leq\frac{s}{4}\sum_{u=0}^{s-1}[f(u+1)-f(u)]^2
 \leq\frac{sZ_s}{2\tau}\mathcal E_s(f,f).
 \label{eq:app-pair-poincare}
\end{equation}
The conditional pair gap is at least
\begin{equation}
 \frac{2\tau}{sZ_s}
 \geq\frac{2\tau}{s(s+1)}
 \geq\frac{\tau}{k^2},
 \qquad 1\leq s\leq k;
 \label{eq:app-pair-gap}
\end{equation}
for $s=1$ it equals $\tau$, and $s=0$ has no conditional variance.

Let $\mathcal E_{\rm HB}$ be the form of
Eq.~\eqref{eq:app-heat-bath-generator}.  Applying
Eq.~\eqref{eq:app-pair-gap} conditionally in every pair and using
Eq.~\eqref{eq:app-mean-field-pair-decomposition} gives, with
$\mathcal V_{ij}(f)$ denoting the expected conditional variance outside
$\{i,j\}$,
\begin{align}
 \mathcal E_{\operatorname{ZRP}(K_k,r_a,k)}(f,f)
 &\geq\frac2k\frac{\tau}{k^2}
 \sum_{i<j}\mathcal V_{ij}(f)
 \\
 &=\frac{2\tau}{k^2}\mathcal E_{\rm HB}(f,f).
 \label{eq:app-zrp-heat-bath-form-comparison}
\end{align}
Combining this with Eq.~\eqref{eq:app-heat-bath-gap} yields
\begin{equation}
 \operatorname{gap}\operatorname{ZRP}(K_k,r_a,k)
 \geq\frac{2\tau}{k^2}\frac1k
 =2\tau k^{-3}.
 \label{eq:app-mean-field-gap-final}
\end{equation}
This displays separately the pair-selection factor $2/k$, the conditional
pair estimate $\tau/k^2$, and the heat-bath gap $1/k$.

Finally, for $k\geq3$,
$1-\cos(2\pi/k)\geq8/k^2$.  Substitution of
Eq.~\eqref{eq:app-mean-field-gap-final} into
Eq.~\eqref{eq:cycle-zrp-reduction} gives
\begin{equation}
 \Delta_{D_n}(a)
 \geq\frac2n\frac8{k^2}\frac{2\tau}{k^3}
 =\frac{32\tau}{nk^5}
 =1024\,n^{-(a+6)},
\end{equation}
because $k=n/2$ and $\tau=n^{-a}$.  This completes the proof of
Proposition~\ref{prop:cycle-uniform-gap}.  Writing
$\Delta_{\rm ZRP}(a)=\operatorname{gap}\operatorname{ZRP}(K_k,r_a,k)$
and also using cyclic-envelope inclusion, the complete protocol-level comparison
is
\begin{multline}
 \Delta_{\rm par}^{\rm cyc}(a)\geq\Delta_{D_n}(a)\\
 \geq\frac2n\left[1-\cos\left(\frac{2\pi}{k}\right)\right]
 \Delta_{\rm ZRP}(a)\\
 \geq1024\,n^{-(a+6)}.
 \label{eq:gap-proof-architecture}
\end{multline}

\section{Numerical protocol and full tables}
\label{app:numerical-protocol}

Inputs are frozen with identifiers and SHA-256 hashes.  Every reported gap
is computed either in the full Hamiltonian or in an explicitly labelled
exact orbit or cyclic restriction.  Values from different operators are
never combined in one unlabeled column.  Numerical degeneracy is accepted
only relative to eigenpair residuals and matrix scale.  Inside a degenerate
manifold, let $P_0$ denote its spectral projector and let
$\Pi_{\alpha\beta}$ project onto the joint irrep labelled by
$(\alpha,\beta)$.  The irrep content is obtained from
\begin{equation}
 \operatorname{rank}(P_0\Pi_{\alpha\beta}P_0),
\end{equation}
not by assigning labels to solver-dependent individual eigenvectors.

In the tables below, $g_0^{\rm full}$ and $g_0^{\rm acc}$ are the ranks of
the ground projectors on $\mathcal H_{\rm addr}$ and $\mathcal K_u$,
respectively.  A partition in square brackets labels the corresponding
$S_k$ irrep.

For sparse full-space calculations, the central class sum of register
transpositions separates every $S_k$ irrep for $k\leq5$.  Acting on a
lowest vector with register permutations constructs its full representation
multiplet; deflation then tests
for further cospectral copies before $g_0^{\rm full}$ is assigned.  For the
grid row, the two-dimensional global ground manifold is the $E$ irrep of
$D_4$, whereas the symmetry-preserving protocol branch is $A_1$.  At the analytic
hopping-cancellation point, the joint-fixed ground rank agrees with the
independently enumerated number of $G_B$-orbits of optimal covers in every
row.

The cyclic basis is constructed independently from the exact potential
shells by Eq.~\eqref{eq:shell-cyclic-iteration}.  Its dimension is stable
for relative singular-value tolerances $10^{-9}$, $10^{-11}$, and
$10^{-13}$ in all eleven rows.  The smallest retained singular value is
$4.4\times10^{-4}$, the largest discarded one is
$1.1\times10^{-15}$, and the largest transport-invariance residual is
$2.3\times10^{-14}$.  The direct residual of
Eq.~\eqref{eq:cancellation-identity} is at most $2.2\times10^{-14}$ across
the audit.  The largest orbit-invariance residual is below
$5\times10^{-13}$, and every
full-space eigenpair and class-sum residual used below is stored in the
machine-readable output.  The shell count $p$ in
Table~\ref{tab:finite-structure} ranges from 9 to 31 and is strictly below
$\dim\mathcal K_u$ in every row; Eq.~\eqref{eq:shell-lower-bound} is thus a
structural floor, not a dimension estimate, on this sample.  Eight rows are
subinstances extracted from the OR-Library Set Cover
benchmarks~\cite{Beasley1990,ORLib}; parent indices, extraction metadata, and
hashes are stored in the instance registry.

\begin{table*}[t]
\centering
\caption{Direct audit of Proposition~\ref{prop:sector-localization-rewrite}. The reported leakage is the worst value in the sector ground manifold; $b_{\Gamma}$ and $b_g$ are the right-hand side of Eq.~\eqref{eq:sector-bound-rewrite} using the exact separation $\Gamma_\gamma$ and the sufficient kinetic-floor separation $g_\gamma=m-\kappa-\eta_\gamma$, respectively.}
\label{tab:sector-localization-audit}
\scriptsize
\setlength{\tabcolsep}{4pt}
\begin{tabular}{llrrrrrrrr}
\hline
instance & sector $\gamma$ & $d_\gamma$ & $\operatorname{rank}P_\gamma$ & $\eta_\gamma$ & $\Gamma_\gamma$ & $\|B_\gamma\|$ & leakage & $b_{\Gamma}$ & $b_g$\\
\hline
grid-2x4 & $[3]\times A1$ & 22 & 2 & 0 & 4.66157 & 1.16709 & 0.00560496 & 0.0589853 & 0.0611906\\
grid-2x4 & $[3]\times A2$ & 8 & 1 & -0.285714 & 4.87141 & 0.377964 & 0.00219248 & 0.00598392 & 0.00601892\\
grid-2x4 & $[3]\times B1$ & 22 & 2 & -0.142857 & 4.73909 & 0.92352 & 0.00346365 & 0.0365861 & 0.0369578\\
grid-2x4 & $[3]\times B2$ & 8 & 1 & 0 & 4.73302 & 0.553283 & 0.0030794 & 0.013481 & 0.014437\\
grid-2x4 & $[3]\times E$ & 60 & 6 & -0.285714 & 4.88189 & 0.889694 & 0.00357428 & 0.0321452 & 0.0324629\\
rnd-b8-t7-p50-s002 & $[3]$ & 120 & 12 & -0.428571 & 5 & 1.19582 & 0 & 0.0541047 & 0.0541047\\
\hline
\end{tabular}
\end{table*}

\begin{table*}[t]
\centering
\caption{Exact structural reduction for the frozen finite instances. All dimensions refer to the physical, unpadded address space; $p$ is the number of potential shells and $d_{\rm miss}=\dim\mathcal H_{\rm sym}-\dim\mathcal K_u$.}
\label{tab:finite-structure}
\scriptsize
\setlength{\tabcolsep}{3pt}
\begin{tabular}{lrrrrrrrr}
\hline
instance & $k$ & $|G_B|$ & $\dim\mathcal H_{\rm addr}$ & $\dim\mathcal H_{\rm sym}$ & $p$ & $\dim\mathcal K_u$ & $d_{\rm miss}$ & cover orbits\\
\hline
grid-2x4 & 3 & 8 & 512 & 22 & 9 & 22 & 0 & 2\\
rnd-b8-t7-p50-s002 & 3 & 1 & 512 & 120 & 9 & 101 & 19 & 12\\
rnd-b8-t8-p42-s014 & 4 & 1 & 4096 & 330 & 15 & 324 & 6 & 7\\
scpe1-b8-t20 & 4 & 1 & 4096 & 330 & 19 & 328 & 2 & 8\\
scpe1-b8-t30 & 5 & 1 & 32768 & 792 & 30 & 792 & 0 & 1\\
scpe2-b8-t20 & 4 & 1 & 4096 & 330 & 19 & 328 & 2 & 3\\
scpe2-b8-t30 & 5 & 1 & 32768 & 792 & 29 & 792 & 0 & 8\\
scpe3-b8-t20 & 3 & 1 & 512 & 120 & 14 & 119 & 1 & 2\\
scpe3-b8-t30 & 5 & 1 & 32768 & 792 & 31 & 792 & 0 & 4\\
scpe4-b8-t29 & 5 & 1 & 32768 & 792 & 30 & 792 & 0 & 1\\
scpe5-b8-t30 & 5 & 1 & 32768 & 792 & 29 & 792 & 0 & 2\\
\hline
\end{tabular}
\end{table*}

\begin{table*}[t]
\centering
\caption{Problem-endpoint spectra reconstructed from the full Hamiltonian and the exact joint-fixed and cyclic restrictions. Here $\Delta$ is the gap above the entire ground manifold, $\delta_{\rm sym}=E_0^{\rm sym}-E_0^{\rm full}$, $\delta_{\rm cyc}=E_0^{\mathcal K}-E_0^{\rm sym}$, and $p_{\rm opt}^{\min}$ is minimized over normalized cyclic ground states. The symmetric and cyclic endpoint ground states are unique in every row; the last column is the joint-fixed ground rank at the hopping-cancellation point.}
\label{tab:finite-endpoints}
\scriptsize
\setlength{\tabcolsep}{3pt}
\begin{tabular}{llrrrrrrrr}
\hline
instance & ground $S_k$ content & $g_0^{\rm full}$ & $\Delta_{\rm full}$ & $\delta_{\rm sym}$ & $\delta_{\rm cyc}$ & $\Delta_{\rm sym}$ & $\Delta_{\rm cyc}$ & $p_{\rm opt}^{\min}$ & $g_0^{\rm sym}(s_c)$\\
\hline
grid-2x4 & $2[3]$ & 2 & 0.000382 & 0.271 & 0 & 0.315 & 0.315 & 0.9944 & 2\\
rnd-b8-t7-p50-s002 & $[3]\oplus2[2,1]\oplus[1,1,1]$ & 6 & 0.0563 & 0 & 0.0581 & 0.0581 & 0.128 & 0.9986 & 12\\
rnd-b8-t8-p42-s014 & $[2,1,1]$ & 3 & 0.000327 & 0.0274 & 0 & 0.0297 & 0.0297 & 0.9955 & 7\\
scpe1-b8-t20 & $[1,1,1,1]$ & 1 & 0.000539 & 0.0103 & 0 & 0.00178 & 0.00178 & 0.9983 & 8\\
scpe1-b8-t30 & $[5]$ & 1 & 0.00591 & 0 & 0 & 4.87 & 4.87 & 0.9949 & 1\\
scpe2-b8-t20 & $[2,1,1]$ & 3 & 0.00137 & 0.00674 & 0 & 0.18 & 0.18 & 0.9966 & 3\\
scpe2-b8-t30 & $[1,1,1,1,1]$ & 1 & 0.00118 & 0.0316 & 0 & 0.0829 & 0.0829 & 0.9982 & 8\\
scpe3-b8-t20 & $[2,1]$ & 2 & 0.000658 & 0.00499 & 0 & 0.256 & 0.256 & 0.9969 & 2\\
scpe3-b8-t30 & $[3,1,1]$ & 6 & 0.000434 & 0.00864 & 0 & 0.247 & 0.247 & 0.9985 & 4\\
scpe4-b8-t29 & $[5]$ & 1 & 0.00608 & 0 & 0 & 4.8 & 4.8 & 0.9945 & 1\\
scpe5-b8-t30 & $[4,1]$ & 4 & 0.00306 & 0.00749 & 0 & 0.242 & 0.242 & 0.9962 & 2\\
\hline
\end{tabular}
\end{table*}

For the parent family, exact $D_n$-orbit quotients for
$n=6,8,\ldots,18$ place the minimum of a 31-point grid at $a=7$.  Endpoint
gaps decrease from $0.166670$ at $n=6$ to $0.0105026$ at $n=18$ and look
approximately quadratic on these sizes.  This is finite-grid evidence, not
a scaling inference; Proposition~\ref{prop:cycle-uniform-gap} supplies the
uniform certificate.

\end{document}